\documentclass{article}

\usepackage{PRIMEarxiv}
\usepackage[utf8]{inputenc}
\usepackage[T1]{fontenc}
\usepackage{hyperref}
\usepackage{url}
\usepackage{booktabs}
\usepackage{amsmath,amssymb}
\usepackage{microtype}
\usepackage{graphicx}
\usepackage{tabularx}
\usepackage{caption}
\usepackage[round,authoryear]{natbib}
\graphicspath{{media/}}
\hypersetup{colorlinks=true,linkcolor=black,citecolor=black,urlcolor=black,
  pdftitle={Abstract Priorities, Concrete Rewards: How Stated Preferences Relate to What People Reward in AI Responses},
  pdfauthor={Zihan Chen, Lei Nico Zheng, Di Zhu},
  pdfkeywords={human-AI interaction, large language models, stated and revealed preferences, preference elicitation, self-insight, human feedback, AI personalization}}
\newcommand{\Npersons}{1,396}
\newcommand{\Nconv}{7,978}
\newcommand{\Nresp}{29,347}
\newcommand{\Nmodels}{21}
\newcommand{\Ncountries}{35}
\newcommand{\MedRange}{52}
\newcommand{\DefTrunc}{12.9}
\newcommand{\DefLeak}{1.3}
\newcommand{\DefRep}{1.8}
\newcommand{\CreMissing}{3.3}
\newcommand{\NBirth}{75}
\newcommand{\MaxConv}{7}
\newcommand{\MedConv}{6}
\newcommand{\NDropShort}{670}
\newcommand{\NStraightSample}{33}
\newcommand{\NIps}{1,363}
\newcommand{\NStraightSurvey}{37}
\newcommand{\NAValUng}{21}
\newcommand{\NADivUng}{19}
\newcommand{\TOneLen}{5.56}
\newcommand{\TOneLenSE}{0.31}
\newcommand{\TOneHelp}{3.69}
\newcommand{\TOneHelpSE}{0.19}
\newcommand{\TOneRef}{$-$2.14}
\newcommand{\TOneRefSE}{0.20}
\newcommand{\TOneStruct}{1.93}
\newcommand{\TOneStructSE}{0.19}
\newcommand{\TOneFlu}{1.58}
\newcommand{\TOneFluSE}{0.22}
\newcommand{\TOneSafe}{0.05}
\newcommand{\TOneSafeSE}{0.16}
\newcommand{\TOneWald}{612}
\newcommand{\RtwoFE}{0.195}
\newcommand{\RtwoAll}{0.256}
\newcommand{\RtwoMatched}{0.226}
\newcommand{\RtwoUnclaimed}{0.231}
\newcommand{\WMLen}{1.84}
\newcommand{\WMLenSE}{0.16}
\newcommand{\WMHelp}{1.12}
\newcommand{\WMRef}{$-$0.81}
\newcommand{\WMFlu}{0.83}
\newcommand{\WMVal}{$-$0.01}
\newcommand{\WMDiv}{0.01}
\newcommand{\WMAcc}{54.8}
\newcommand{\WMAccSE}{0.4}
\newcommand{\WMPairs}{17,908}
\newcommand{\CtxLenPrompt}{1.38}
\newcommand{\CtxLenContro}{1.68}
\newcommand{\CtxRefContro}{1.18}
\newcommand{\CtxHelpContro}{$-$1.43}
\newcommand{\CVFlu}{0.09}
\newcommand{\CVDiv}{0.08}
\newcommand{\CVLenFact}{0.09}
\newcommand{\CVLenDiv}{0.09}
\newcommand{\CVStructHelp}{0.07}
\newcommand{\CVStructCre}{0.06}
\newcommand{\CVRefCre}{$-$0.06}
\newcommand{\CVRefHelp}{$-$0.05}
\newcommand{\CVFactFact}{0.01}
\newcommand{\CVHelpHelp}{0.01}
\newcommand{\CVSafeSafe}{$-$0.01}
\newcommand{\CVN}{7,041}
\newcommand{\CVLvLenFact}{0.16}
\newcommand{\CVLvLenDiv}{0.15}
\newcommand{\CVLvRefHelp}{$-$0.12}
\newcommand{\CVLvRefCre}{$-$0.11}
\newcommand{\CVLvHelpHelp}{0.07}
\newcommand{\CVLvValVal}{0.07}
\newcommand{\CVNTopNonElicited}{5}
\newcommand{\CVNTopNonElicitedLv}{5}
\newcommand{\TTwoGamma}{0.09}
\newcommand{\TTwoSE}{0.08}
\newcommand{\TTwoP}{0.30}
\newcommand{\TTwoCIlo}{$-$0.08}
\newcommand{\TTwoCIhi}{0.25}
\newcommand{\TTwoWald}{10.72}
\newcommand{\TTwoWaldP}{0.15}
\newcommand{\IpsHelp}{0.69}
\newcommand{\IpsHelpP}{0.02}
\newcommand{\IpsHelpHolm}{0.15}
\newcommand{\RawHelp}{0.50}
\newcommand{\RawHelpSE}{0.17}
\newcommand{\RawHelpHolm}{0.025}
\newcommand{\TTwoPermP}{0.17}
\newcommand{\RobMin}{$-$0.04}
\newcommand{\RobMax}{0.14}
\newcommand{\LomoMin}{0.02}
\newcommand{\LomoMax}{0.11}
\newcommand{\AltWaldP}{0.02}
\newcommand{\LevProx}{0.00}
\newcommand{\LevProxSE}{0.09}
\newcommand{\LevProxP}{$p=0.99$}
\newcommand{\LevSepProx}{0.08}
\newcommand{\LevSepProxP}{$p=0.31$}
\newcommand{\LevLens}{0.33}
\newcommand{\LevLensSE}{0.14}
\newcommand{\LevLensP}{$p=0.01$}
\newcommand{\LevSepLens}{0.42}
\newcommand{\LevSepLensP}{$p<0.001$}
\newcommand{\LevBootLensLo}{$-$0.10}
\newcommand{\LevBootLensHi}{0.56}
\newcommand{\LevProf}{0.85}
\newcommand{\LevProfSE}{0.23}
\newcommand{\LevProfP}{$p<0.001$}
\newcommand{\LevSepProf}{0.93}
\newcommand{\LevSepProfP}{$p<0.001$}
\newcommand{\LevBootProfLo}{0.15}
\newcommand{\LevBootProfHi}{1.13}
\newcommand{\DiffProfProx}{0.85}
\newcommand{\DiffProfProxSE}{0.25}
\newcommand{\DiffProfProxHolm}{$p_{\mathrm{Holm}}=0.002$}
\newcommand{\DiffProfProxBootLo}{0.16}
\newcommand{\DiffProfProxBootHi}{1.12}
\newcommand{\DiffProfLens}{0.51}
\newcommand{\DiffProfLensHolm}{$p_{\mathrm{Holm}}=0.13$}
\newcommand{\DiffLensProx}{0.33}
\newcommand{\DiffLensProxHolm}{$p_{\mathrm{Holm}}=0.13$}
\newcommand{\RepDiffMin}{0.57}
\newcommand{\RepShareProfGtProx}{100}
\newcommand{\RepN}{50}
\newcommand{\LensHeldLo}{0.01}
\newcommand{\LensHeldHi}{0.12}
\newcommand{\ProfRelLo}{0.35}
\newcommand{\ProfRelHi}{0.83}
\newcommand{\QualRel}{0.69}
\newcommand{\SpecMatched}{0.93}
\newcommand{\SpecDev}{0.29}
\newcommand{\SpecDevSE}{0.08}
\newcommand{\SpecDevP}{$p<0.001$}
\newcommand{\SpecMisMean}{$-$0.06}
\newcommand{\SpecMisMax}{0.28}
\newcommand{\ConsDis}{0.93}
\newcommand{\ConsDisSE}{0.21}
\newcommand{\ConsCorr}{0.92}
\newcommand{\CrossMinHolm}{0.37}
\newcommand{\ExpMinP}{0.25}
\newcommand{\ConcShortPart}{7}
\newcommand{\ConcLongPart}{24}
\newcommand{\ConcBrevPraisePart}{12}
\newcommand{\ConcDetPraisePart}{32}
\newcommand{\ConcFmtPart}{12}
\newcommand{\ConcRefPart}{3}
\newcommand{\AudShortPrec}{0.77}
\newcommand{\AudShortLo}{0.57}
\newcommand{\AudShortHi}{0.90}
\newcommand{\AudShortN}{22}
\newcommand{\AudLongPrec}{0.75}
\newcommand{\AudLongLo}{0.60}
\newcommand{\AudLongHi}{0.86}
\newcommand{\AudLongN}{40}
\newcommand{\AudFmtPrec}{0.80}
\newcommand{\AudFmtLo}{0.63}
\newcommand{\AudFmtHi}{0.90}
\newcommand{\AudShortRecall}{0.22}
\newcommand{\AudShortRecallLo}{0.12}
\newcommand{\AudShortRecallHi}{0.56}
\newcommand{\AudLongRecall}{0.18}
\newcommand{\AudLongRecallLo}{0.11}
\newcommand{\AudLongRecallHi}{0.38}
\newcommand{\AudItems}{222}
\newcommand{\AudPoolN}{7,518}
\newcommand{\AudAIAgree}{96}
\newcommand{\AudAIKappa}{0.91}
\newcommand{\AudAIN}{142}
\newcommand{\AudAITexts}{102}
\newcommand{\ConcLenT}{$-$6.16}
\newcommand{\ConcLenTSE}{1.54}
\newcommand{\ConcLenTHolm}{$p_{\mathrm{Holm}}<0.001$}
\newcommand{\ConcLenTSD}{$-$0.78}
\newcommand{\ConcLenE}{$-$3.50}
\newcommand{\ConcLenESE}{1.05}
\newcommand{\ConcLenEHolm}{$p_{\mathrm{Holm}}=0.002$}
\newcommand{\ConcLenC}{$-$0.39}
\newcommand{\ConcLenCP}{$p=0.54$}
\newcommand{\ConcFmtT}{2.93}
\newcommand{\ConcFmtTHolm}{0.18}
\newcommand{\RawValLen}{$-$0.78}
\newcommand{\RawValLenP}{$p<0.001$}
\newcommand{\RefAttrIpsP}{0.09}
\newcommand{\CommonRawPriP}{0.006}
\newcommand{\CommonIpsConc}{$-$6.38}
\newcommand{\CommonIpsConcSE}{1.53}
\newcommand{\CommonIpsConcP}{$p<0.001$}
\newcommand{\CommonIpsPriP}{0.171}
\newcommand{\CommonIpsPriDf}{6}
\newcommand{\RsnMoneCue}{0.031}
\newcommand{\RsnMoneLen}{0.045}
\newcommand{\RsnMtwoOwn}{0.55}
\newcommand{\RsnMtwoCue}{0.013}
\newcommand{\RsnMtwoLen}{0.026}
\newcommand{\RsnMthreeOthers}{0.12}
\newcommand{\SpecOwnLo}{0.20}
\newcommand{\SpecOwnHi}{0.58}
\newcommand{\SpecCrossLo}{0.00}
\newcommand{\SpecCrossHi}{0.07}
\newcommand{\TauPLen}{0.17}
\newcommand{\LRLen}{103}
\newcommand{\BetaStdLen}{0.23}
\newcommand{\TauPRef}{0.12}
\newcommand{\TauPFlu}{0.10}
\newcommand{\TauPStruct}{0.08}
\newcommand{\TauPFact}{0.09}
\newcommand{\LRRef}{30}
\newcommand{\NSigPerson}{6}
\newcommand{\NSigPersonHolm}{5}
\newcommand{\TauPLenCtx}{0.17}
\newcommand{\LRLenCtx}{92}
\newcommand{\NSigPersonCtx}{4}
\newcommand{\CalNSim}{100}
\newcommand{\CalRejLo}{2}
\newcommand{\CalRejHi}{7}
\newcommand{\CalLRNullMax}{5.7}
\newcommand{\CalLRObsMin}{15}
\newcommand{\CalRejHet}{9}
\newcommand{\CalLRHetMax}{7.8}
\newcommand{\CalPowerCount}{50}
\newcommand{\CalTauTrue}{0.171}
\newcommand{\CalTauHat}{0.172}
\newcommand{\CalNNon}{50}
\newcommand{\SHrMean}{0.021}
\newcommand{\SHrLo}{$-$0.03}
\newcommand{\SHrHi}{0.05}
\newcommand{\MomLenLo}{$-$10.2}
\newcommand{\MomLenHi}{19.9}
\newcommand{\MomNExcl}{2}
\newcommand{\SimFitLo}{0.006}
\newcommand{\SimFitHi}{0.044}
\newcommand{\SimFitLen}{0.044}
\newcommand{\SimTwoHi}{0.133}
\newcommand{\SimNullAbsMax}{0.004}
\newcommand{\SimMCSEMax}{0.005}
\newcommand{\SimNRep}{100}
\newcommand{\RelReasonRLo}{0.55}
\newcommand{\RelReasonRHi}{0.76}
\newcommand{\RelReasonIpsRLo}{0.29}
\newcommand{\RelReasonIpsRHi}{0.54}
\newcommand{\BaseAcc}{65.8}
\newcommand{\NpredEligible}{1,309}
\newcommand{\NpredPersons}{1,307}
\newcommand{\GSlider}{$-$0.23}
\newcommand{\GSliderLo}{$-$0.45}
\newcommand{\GSliderHi}{0.00}
\newcommand{\GProf}{0.02}
\newcommand{\GProfLo}{$-$0.11}
\newcommand{\GProfHi}{0.17}
\newcommand{\GProfMSE}{$-$1.3}
\newcommand{\GProfMSELo}{$-$2.2}
\newcommand{\GProfMSEHi}{$-$0.3}
\newcommand{\GNbr}{$-$0.11}
\newcommand{\GDemo}{$-$0.55}
\newcommand{\GDemoLo}{$-$0.85}
\newcommand{\GDemoHi}{$-$0.26}
\newcommand{\GReason}{$-$0.17}
\newcommand{\GConc}{0.13}
\newcommand{\GConcLo}{0.02}
\newcommand{\GConcHi}{0.24}
\newcommand{\GFTWish}{0.06}
\newcommand{\GFTWishLo}{$-$0.01}
\newcommand{\GFTWishHi}{0.13}
\newcommand{\GOwnCue}{$-$0.02}
\newcommand{\GOwnCueMSE}{$-$2.1}
\newcommand{\GOwnCueMSELo}{$-$3.7}
\newcommand{\GOwnCueMSEHi}{$-$0.6}
\newcommand{\GOwnModel}{0.13}
\newcommand{\GOwnModelLo}{$-$0.04}
\newcommand{\GOwnModelHi}{0.30}
\newcommand{\GGBM}{$-$0.07}
\newcommand{\GMax}{0.13}
\newcommand{\GHiMax}{0.30}
\newcommand{\NAdditions}{12}
\newcommand{\GBMPool}{1.11}
\newcommand{\GBMPoolLo}{0.73}
\newcommand{\GBMPoolHi}{1.52}
\newcommand{\LinPairPool}{0.35}
\newcommand{\LinPairPoolLo}{0.12}
\newcommand{\LinPairPoolHi}{0.62}
\newcommand{\GBMvsLin}{0.76}
\newcommand{\GBMvsLinLo}{0.43}
\newcommand{\GBMvsLinHi}{1.10}
\newcommand{\WANSig}{1}
\newcommand{\WANTot}{6}
\newcommand{\WAMax}{0.36}
\newcommand{\WAMaxLo}{0.10}
\newcommand{\WAMaxHi}{0.63}
\newcommand{\PDConcSlid}{0.36}
\newcommand{\PDConcSlidLo}{0.12}
\newcommand{\PDConcSlidHi}{0.60}
\newcommand{\PDConcProf}{0.11}
\newcommand{\PDConcProfLo}{$-$0.06}
\newcommand{\PDConcProfHi}{0.26}
\newcommand{\PDBothConc}{$-$0.33}
\newcommand{\PDBothConcLo}{$-$0.57}
\newcommand{\PDBothConcHi}{$-$0.11}
\newcommand{\ChModelFive}{0.32}
\newcommand{\ChModelFiveLo}{$-$0.17}
\newcommand{\ChModelFiveHi}{0.79}
\newcommand{\TrModel}{0.04}
\newcommand{\TrModelLo}{$-$0.09}
\newcommand{\TrModelHi}{0.16}
\newcommand{\OrderMax}{1.4}

\title{Abstract Priorities, Concrete Rewards: How Stated Preferences Relate to What People Reward in AI Responses
\thanks{Working paper. The code, scripts, and analysis data are available from the authors upon request and will be released after the review process.}}

\author{
  Zihan Chen \\
  Stevens Institute of Technology \\
  Hoboken, NJ, USA \\
  \texttt{zchen61@stevens.edu} \\
  \And
  Lei Nico Zheng \\
  University of Massachusetts Boston \\
  Boston, MA, USA \\
  \texttt{lei.zheng@umb.edu} \\
  \And
  Di Zhu \\
  Stevens Institute of Technology \\
  Hoboken, NJ, USA \\
  \texttt{dzhu1@stevens.edu} \\
}

\begin{document}
\maketitle

\begin{abstract}
AI assistants built on large language models (LLMs) are trained on human feedback and increasingly personalized to users' stated preferences. Yet how closely these preferences reflect what people reward in responses remains unclear. We compare stated and revealed preferences in the public PRISM dataset, in which \Npersons{} participants reported how much seven response attributes mattered to them and then rated \Nresp{} responses from \Nmodels{} LLMs. Among responses to the same prompt, features the survey never asked about, such as response length, formatting and refusal language, were at least as strongly associated with ratings as text proxies of the surveyed attributes. Whether stated priorities matched ratings depended on how attributes were measured: participants who prioritized an attribute favoured responses from models that others perceived as strong on it, but not responses that scored higher on text proxies for it. Written feedback was more telling: participants who asked for shorter answers, even in earlier conversations, gave length less weight in their ratings, and those asking for longer answers gave it more. Their relative priorities over the surveyed attributes, however, did not predict this weight. Reasons given for choices reflected perceived attributes more than measured features. Participants also differed stably in how much they weighted several concrete features, especially length, but with a median of \MedConv{} conversations per participant, these weights were estimated poorly. The largest gain from personal information in predicting held-out choices was \GMax{} percentage points, compared with \GBMPool{} for a stronger pooled model. Stated priorities thus track perceived attributes of AI responses but say little about the concrete features that ratings reward, a distinction that matters for preference elicitation and AI personalization.
\end{abstract}


\section{Introduction}

Conversational AI systems are increasingly tailored to what users say they want. Settings pages and custom instructions invite users to describe their preferences, typically in terms of broad attributes: accurate, safe, concise, creative. Research on pluralistic and personalized alignment proposes to condition models on profiles of this kind \citep{sorensen2024roadmap,kirk2024benefits,poddar2024personalizing,ryan2025synthesizeme}, and preference-based training relies on ratings of concrete responses \citep{ouyang2022training,casper2023open}. Both practices raise a psychological question: how do the priorities people state relate to what they reward when they rate the responses in front of them?

Research on judgment gives reasons to expect an imperfect correspondence. The weights people report for different pieces of information correspond only partly to the weights recovered from their judgments \citep{slovic1971comparison,reilly1992assessment,karelaia2008determinants}; explanations of choices often reflect plausible theories of what should have mattered \citep{nisbett1977telling}; and stated attitudes diverge from behaviour in many domains, including privacy \citep{kokolakis2017privacy,barth2017privacy}. A weak relation can arise for different reasons, however. People may lack insight into what drives their evaluations. The measures that link statements to behaviour may not correspond to how people represent the attributes they name. Or preferences may depend on context, so that a single summary cannot capture them. These explanations have different consequences for the design of preference elicitation, and they are rarely distinguished.

We examine whether the correspondence depends on how the attributes named in a preference report are measured. Stated priorities are abstract: they name attributes such as ``factual and informative''. The responses people rate differ in concrete features such as length, formatting, or refusal language. In the lens model of judgment, a reader infers qualities such as informativeness from the cues a response provides \citep{hammond1955probabilistic,karelaia2008determinants}. Construal-level theory links psychological distance to the relative emphasis on abstract and concrete features \citep{trope2010construal}. If stated priorities are formulated at the level of perceived attributes, they may align with behaviour measured at that level while saying little about the concrete cues that accompany the perceptions. Here, alignment means that participants who give an attribute greater priority also reward higher measured values of that attribute more strongly when they rate responses.

We study this question with PRISM \citep{kirk2024prism}, a public dataset in which participants first rated how important seven attributes of AI responses were to them and then held several conversations. In each conversation, up to four models answered the same opening prompt, and the participant rated every response and reported why they chose the one they preferred. Comparisons within a conversation hold the prompt and participant constant, reducing variation due to differences in requests and raters. The PRISM authors noted that per-conversation reasons correlate only weakly with stated preferences \citep{kirk2024prism}. Two recent studies are closest to ours. PRISM-X re-recruited 530 PRISM participants two years later and found that fine-tuning on an individual's preference data yields only marginal gains over pooled preferences \citep{kirk2026prismx}. LiteraryTaste linked 60 readers' stated reading preferences to their choices over 100 pairs of creative texts each and found that stated preferences had limited utility for modelling revealed preferences \citep{chung2025literarytaste}. These studies establish the relevance of the stated--revealed preference gap. We investigate how its interpretation changes when the same stated attributes are linked to response features, perceived attributes and reasons for choosing a response, across several conversation types.

The analysis follows the path from response evaluation to personalization. We first identify the features associated with ratings and with perceived attributes. We then test whether stated priorities predict the weights attached to those attributes under different measurements, and compare what abstract priorities, concrete feedback and reported reasons reveal. Finally, we ask whether individual feature weights are stable, recoverable from short histories and useful for predicting held-out choices. Across these analyses, priorities align more clearly with aggregate attribute profiles, built from other participants' perceptions, than with designed text proxies, concrete requests track the weight on length, and stable individual differences coexist with small predictive gains from personal information.

The paper makes three contributions. First, it shows that the apparent correspondence between stated priorities and response ratings depends on how the attributes are operationalized. This makes measurement an integral part of interpreting self-insight in human--AI evaluation. Second, it distinguishes the information conveyed by abstract priorities, concrete requests and choice-specific reasons, identifying an association between requests about length and the weight attached to length. Third, it separates stable individual differences from their recoverability and predictive value, using variance-component models, recovery simulations and held-out comparisons against a common pooled baseline. Together, these contributions connect a psychological question about preference reports to a practical question about personalization.

\section{Theoretical background}

The theoretical argument links three elements of evaluation: the observable cues in a response, the attributes a reader perceives and the preferences or reasons the reader reports. Judgment analysis explains how cues enter evaluations; research on self-report and construal explains why a report may correspond more closely to one element than another. Work on human feedback and personalization provides the setting in which these distinctions matter.

\subsection{Cues, perceived attributes and judgment}

Judgment-analysis research represents an evaluation as a policy linking the cues of an object to the judgment it receives \citep{hammond1955probabilistic,karelaia2008determinants}. In the lens model, a judge cannot observe the qualities of an object directly and infers them from proximal cues, which may relate to the qualities in complex, many-to-many ways. Policy capturing estimates cue weights from repeated judgments of objects whose cues vary, and self-insight is assessed by comparing captured weights with the weights judges report \citep{karren2002review,aimansmith2002conducting}. Self-insight in such studies is real but partial and depends on how it is measured \citep{slovic1971comparison,reilly1992assessment}. A low correspondence between reported and captured weights can therefore reflect missing insight, inconsistent judgment, or a mismatch between the cues the analyst measures and the attributes the judge reports.

\subsection{What reports reveal}

\citet{nisbett1977telling} argued that people have limited introspective access to the causes of their judgments and that reports of those causes draw on a priori theories, so they are accurate when the theories are correct and inaccurate when the actual influences are non-obvious. Global impressions can also colour ratings of specific attributes \citep{thorndike1920constant,nisbett1977halo}, and people can explain choices they did not make \citep{johansson2005failure}. Reports can nevertheless describe the content of an evaluation. When a report and the behaviour it describes are framed at the same level, a reason such as ``I chose this response because it was helpful'' may accurately summarize a perception that accompanied the choice. This distinction separates reporting a perceived attribute from identifying the cues associated with that perception.

\subsection{Levels of description}

Preferences are often constructed at the moment of elicitation \citep{slovic1995construction}, and the representation used depends on the task. Construal-level research links greater psychological distance to more abstract representations and to changes in the features that influence evaluation \citep{trope2010construal}. A preference survey asks people to evaluate attributes in general; a rating task asks them to evaluate a particular response. Processing-fluency research adds that concrete properties of a message, such as how easy it is to read, shape impressions of its quality independently of its substance \citep{alter2009uniting}. Together these ideas suggest that stated priorities will correspond more closely to perceived attributes than to the concrete features that accompany the perceptions. This motivates a comparison of text-based and perception-based representations of the same attributes, which tests whether alignment is sensitive to how the attributes are measured. The Discussion then weighs level of description against the other explanations in Table~\ref{tab:explanations}.

\subsection{Human feedback and personalization}

Ratings of language-model outputs reward features other than those raters are asked to judge. Preference models and automatic evaluators trained or prompted to mimic human judgments favour longer responses \citep{singhal2023long,dubois2024length}; assertive responses are judged more factual than they are \citep{hosking2024human}; responses that agree with the user are preferred \citep{sharma2024towards}; and human and model judges share several such biases \citep{li2024dissecting,zheng2023judging,chiang2024chatbot}. Some variation in feedback may reflect judgments that change across occasions \citep{ghafouri2026rlhf,keswani2025moral}. Work on personalization argues that people differ in what they want from AI \citep{sorensen2024roadmap,kirk2024benefits,santurkar2023whose}, while studies of prompting show that people find it hard to articulate what they want \citep{zamfirescu2023johnny,subramonyam2024bridging}. Stated preferences have been compared with revealed choices over generated creative texts \citep{chung2025literarytaste}, and personalized fine-tuning has been evaluated with re-recruited PRISM participants \citep{kirk2026prismx}. What remains open is how attribute-specific priorities relate to the features and perceptions that accompany people's own ratings, and whether the answer changes when the same attributes are operationalized in different ways.

\subsection{Competing explanations and the research questions}

Table~\ref{tab:explanations} summarizes four potentially overlapping explanations of a weak or measure-dependent link between stated priorities and ratings, together with the evidence that bears on each.

\begin{table}[t]
\centering
\caption{Explanations of a weak or measure-dependent link between stated priorities and ratings, and the evidence used here.}
\label{tab:explanations}
\small
\begin{tabularx}{\textwidth}{p{3.0cm} X X}
\toprule
Explanation & Prediction & Evidence in this study \\
\midrule
Limited self-insight & Weak alignment persists when attributes are measured in ways that match participants' perceptions and individual policies are reliable & Alignment by representation (Section~\ref{sec:levels}); reliability and recoverability of individual weights (Section~\ref{sec:indiv}) \\
Measurement mismatch & Alignment strengthens as attribute measures move closer to how participants perceive the attributes & Perceptual correspondence of text proxies (Section~\ref{sec:cv}); proxies versus lens composites versus perceived profiles (Section~\ref{sec:levels}) \\
Aggregation, reliability and general quality & Alignment with aggregate profiles reflects averaging, a shared quality dimension or selection into chosen responses rather than attribute-specific perceptions & Attribute-specific profile deviations, mismatched pairings and profile reliability (Section~\ref{sec:levels}) \\
Context dependence & Feature weights vary with the request in systematic ways that a single summary cannot represent & Context interactions (Section~\ref{sec:assoc}); occasion versus person components (Section~\ref{sec:indiv}) \\
\bottomrule
\end{tabularx}
\end{table}

The analysis combines planned hypotheses with exploratory questions. We retain the hypothesis numbers from the analysis plan; the planned experience hypothesis (H3) is treated as exploratory (Section~\ref{sec:plan}). H1 predicts that stated importance moderates the weight on the text proxy designed for each attribute, with the mean of the seven moderation coefficients as the primary estimand. H2 predicts that features omitted from the preference survey are associated with ratings, conditional on prompt and model. H4 predicts that reported reasons are more strongly associated with perceived performance than with measured feature differences. H5 predicts that information from a participant's other conversations improves prediction more than stated priorities do, assessed by a direct comparison.

The exploratory questions examine the explanations in Table~\ref{tab:explanations}. RQ1 asks whether alignment depends on how attributes are operationalized. RQ2 asks whether individual feature weights are stable and recoverable from short histories. RQ3 asks whether spontaneous concrete feedback tracks the weights on corresponding features. RQ4 asks whether experience with language models changes alignment.

\section{Method}

PRISM links pre-conversation priorities to repeated response ratings and post-conversation feedback from the same participants. This structure supports both within-prompt comparisons of responses and tests of whether information about a participant improves prediction across conversations.

\subsection{Data}

PRISM \citep{kirk2024prism} was collected in late 2023 from crowdworkers recruited for geographic and demographic diversity, including census-representative samples for the United States and the United Kingdom. The original study received ethics board approval and participants gave informed consent \citep{kirk2024prism}; this secondary analysis uses the public, de-identified release. Participants first completed a survey covering demographics, familiarity with and use of language models, a free-text self-description, a free-text description of what they want an AI to do, and the importance of eight response attributes on 0--100 visual analogue scales. Each item read ``It is important that an AI language model\ldots'' followed by a statement such as ``\ldots produces factual and informative responses''. Item order was randomized for each participant, so each attribute appears approximately equally often in every position. We refer to these importance ratings as stated priorities.

Participants then held a median of \MedConv{} conversations (at most \MaxConv{}) of three types: unguided, about their values, or about a controversial topic. In the first turn of each conversation, up to four different models, drawn from \Nmodels{}, answered the same prompt, and the participant rated each response on a 1--100 scale. After the conversation ended, participants were asked to ``consider your first message and the top-rated response'' and rated its performance on seven attributes (``This response\ldots''), and then the importance of each attribute in choosing it over the other responses (``I chose this response\ldots''), with an N/A option \citep{kirk2024prism}. These ratings were therefore given after the full conversation and refer to the chosen first-turn response.

The analysis sample comprises all rated first-turn responses with at least three words in conversations with at least two such responses: \Nresp{} responses in \Nconv{} conversations from \Npersons{} participants born in \NBirth{} countries and residing in \Ncountries{} (Table~\ref{tab:sample}; sample flow in Table~\ref{tab:flow}). We excluded \NDropShort{} responses with fewer than three words, mostly one-word outputs of a single model. Ratings varied widely within conversations (median range \MedRange{} points), which is the variation the within-prompt design uses.

\begin{table}[t]
\centering
\caption{Analysis sample (Panel A) and stated importance, reasons and perceived performance by attribute (Panel B). Stated importance uses a 0--100 scale; reasons and perceived performance use 1--100 scales and refer to the chosen first-turn response.}
\label{tab:sample}
\small
\textbf{Panel A. Sample}\\[2pt]
\begin{tabular}{lr}
\toprule
Quantity & Value \\
\midrule
Participants & 1,396 \\
Countries of residence & 35 \\
Conversations & 7,978 \\
\quad unguided & 3,103 \\
\quad values guided & 2,447 \\
\quad controversy guided & 2,428 \\
First-turn rated responses & 29,347 \\
Responses per conversation (mean) & 3.7 \\
Conversations per participant (mean) & 5.7 \\
Models & 21 \\
Median within-conversation score range & 52 \\
Median response length (words) & 57 \\
Familiarity with language models: not / somewhat / very (\%) & 10 / 61 / 28 \\
Female (\%) & 48.5 \\
\bottomrule
\end{tabular}

\vspace{8pt}
\textbf{Panel B. Stated, reported and perceived ratings by attribute}\\[2pt]
\begin{tabular}{lrrrr}
\toprule
Attribute & Stated (mean) & Stated (SD) & Reason (mean) & Perceived performance (mean) \\
\midrule
Values & 54.2 & 26.3 & 67.0 & 74.2 \\
Creativity & 69.8 & 22.2 & 62.1 & 63.7 \\
Fluency & 86.8 & 16.2 & 82.6 & 84.4 \\
Factuality & 88.7 & 16.2 & 79.4 & 79.2 \\
Diversity & 76.0 & 19.8 & 66.0 & 68.8 \\
Safety & 80.5 & 25.1 & 72.1 & 85.1 \\
Helpfulness & 89.6 & 14.3 & 82.6 & 81.6 \\
\bottomrule
\end{tabular}
\end{table}

\subsection{Measures}
\label{sec:measures}

\textbf{Stated priorities.} For each participant we ipsatized the seven importance ratings that also appear in the conversation feedback (subtracting the participant's mean and dividing by their standard deviation), to distinguish relative priorities when several importance ratings cluster near the ceiling. Ipsatized importance describes which attributes a participant considers relatively more important; a coefficient on it is expressed per within-person standard deviation. Of the \Npersons{} analysed participants, \NStraightSample{} gave identical ratings to all seven attributes (\NStraightSurvey{} in the full survey) and have no relative profile; they are excluded from analyses that use ipsatized importance (\NIps{} participants remain). Raw importance, standardized across participants, addresses a different question (absolute rather than relative importance) and is reported alongside.

\textbf{Reasons and perceived performance.} Reasons are the ``I chose this response\ldots'' ratings and perceived performance the ``This response\ldots'' ratings. N/A responses are treated as missing; they were most common for values and diversity in unguided conversations (\NAValUng\% and \NADivUng\%; Table~\ref{tab:na}).

\textbf{Response features.} Twelve features were computed from the text of each response, without reference to any rating (Table~\ref{tab:cues}). Seven were designed to match the wording of the seven attribute items (text proxies); five describe properties that the survey did not elicit (non-elicited features). The feature names describe their operations, and the perceptual-correspondence analysis examines how each relates to participants' perceptions. Semantic similarity uses the all-MiniLM-L6-v2 sentence-embedding model \citep{reimers2019sentence} run locally with sentence-transformers 6.1.0; positive sentiment uses a rule-based sentiment model \citep{hutto2014vader}; lexical diversity uses the measure of textual lexical diversity (MTLD; \citealp{mccarthy2010mtld}). Each feature was demeaned within conversation and then divided by the standard deviation of the demeaned values pooled over all responses, so its coefficient is in rating points per within-conversation standard deviation.

\begin{table}[t]
\centering
\caption{Response features, named by their operations. The middle column gives the survey item each designed feature was meant to approximate. All lexicons are listed in Appendix~\ref{app:lexicons}.}
\label{tab:cues}
\small
\begin{tabularx}{\textwidth}{l X X}
\toprule
Feature & Item it was designed to match (``It is important that an AI language model\ldots'') & Operation \\
\midrule
Coherence & \ldots produces responses that are well-written and coherent & minus the number of defects: truncation, fabricated continuation of the dialogue, trigram repetition above 10\% \\
Informational density & \ldots produces factual and informative responses & numbers and capitalized non-initial words per 100 words \\
Viewpoint markers & \ldots summarises multiple viewpoints or different worldviews & log(1 + viewpoint phrases) \\
Caution markers & \ldots produces responses that are safe and do not risk harm & log(1 + caution and caveat phrases) \\
Relevance to prompt & \ldots produces responses that are helpful and relevant to my requests & embedding similarity of response and prompt \\
Similarity to own self-description & \ldots reflects my values or cultural perspectives & embedding similarity of response and the participant's self-description \\
Lexical diversity & \ldots produces responses that are creative and inspiring & MTLD \\
\midrule
Length & not elicited & log word count \\
Lists and headings & not elicited & log(1 + list, heading and bold lines) \\
Positive sentiment & not elicited & sentiment compound score \\
Validation phrases & not elicited & log(1 + praise or validation phrases) \\
Refusal or AI-disclaimer language & not elicited & log(1 + refusal or disclaimer phrases) \\
\bottomrule
\end{tabularx}
\end{table}

\textbf{Attribute representations.} We represent each attribute in three ways. The first is the designed text proxy in Table~\ref{tab:cues}. The second is a lens composite: a weighted sum of all twelve features, with weights given by how strongly each feature's chosen-minus-other difference predicts participants' perceived performance on that attribute. The third is an aggregate attribute profile of the model that produced the response: other participants' mean perceived performance of that model on the attribute, after removing from each rater's perceptions the part predicted by the rater's own stated priorities (echo correction). Lens weights, echo corrections and profiles were all estimated on a random half of participants and applied to the other half. Profiles summarize other participants' perceptions; the rating participant never saw them. The three representations therefore comprise a single designed feature, a combination of features fitted to perceptions, and an aggregate of reported perceptions. The first two vary between responses; the third summarizes the producing model across chosen responses.

\textbf{Concrete feedback.} Participants' open feedback on each conversation (median 25 words) was coded with direction-aware rules applied clause by clause, with negation handling (Appendix~\ref{app:lexicons}); we call the resulting codes concrete feedback. The rules separate requests for change (shorter or more concise responses; longer or more detailed ones) from praise of brevity or of detail, and code mentions of formatting and complaints about refusal or evasion. They were developed on samples of feedback that were then set aside. An independent coder, who saw neither the rules nor their output, coded a fresh stratified sample of \AudItems{} texts from the remaining \AudPoolN{} conversations, answering six yes/no questions per text from a written codebook (Appendix~\ref{app:lexicons}; Table~\ref{tab:audit}). Against this coding, precision was \AudShortPrec{} for requests for shorter responses (95\% CI \AudShortLo{} to \AudShortHi{}; all \AudShortN{} flagged conversations), \AudLongPrec{} for requests for longer ones (\AudLongLo{} to \AudLongHi{}; \AudLongN{} sampled) and \AudFmtPrec{} for formatting mentions (\AudFmtLo{} to \AudFmtHi{}). Projecting the requests that the coder identified among sampled unflagged texts onto the pool gave estimated recall of \AudShortRecall{} (\AudShortRecallLo{} to \AudShortRecallHi{}) and \AudLongRecall{} (\AudLongRecallLo{} to \AudLongRecallHi{}): the rules capture explicit requests and miss many indirect ones, such as requests for more examples or options. The rules thus detect a subset of requests, together with some false positives, and the prevalences reported below describe detected feedback. For the \AudAITexts{} sampled texts that had also been labelled in an earlier AI-assisted audit, the coder's answers to the two questions about requests for shorter or longer responses agreed with the earlier labels on \AudAIAgree\% of \AudAIN{} judgements ($\kappa = \AudAIKappa$). The same coding function is used in every analysis.

\subsection{Estimation}

\textbf{Conditional associations.} For participant $i$, conversation $c$ and response $j$, the rating $s_{icj}$ is modelled as
\begin{equation}
s_{icj} = \alpha_c + \sum_{k} \beta_k x_{cjk} + \mu_{m(j)} + \pi_{p(j)} + \varepsilon_{icj},
\label{eq:t1}
\end{equation}
with conversation fixed effects $\alpha_c$, standardized features $x_{cjk}$, fixed effects for the producing model $\mu_{m(j)}$ and display position $\pi_{p(j)}$, and standard errors clustered by participant \citep{cameron2015practitioner}. Conversation fixed effects remove influences shared by the responses in a conversation, so each coefficient is the association of a feature with the rating, conditional on the other features, among responses to the same prompt. We re-estimated the model in later turns, where participants rated two samples from the same model, and added interactions with conversation type and prompt length.

\textbf{Perceptual correspondence.} To see how each feature relates to the perception it was meant to approximate, we regressed, for each attribute, the participant's perceived performance of the chosen response on the chosen-minus-other differences in all twelve features, within participant. The estimand is the partial within-person association between a relative feature contrast and an absolute perception of the chosen response. We also report the bivariate associations and a version that uses the chosen response's own feature levels, controlling for conversation type and prompt length.

\textbf{Moderation by stated priorities.} Each attribute weight was allowed to vary with the participant's stated importance, $\beta_k + \gamma_k S_{ik}$. The primary estimand is the mean of the seven $\gamma_k$; we also report each $\gamma_k$ with Holm correction and a joint seven-degree-of-freedom Wald test. As a profile-reassignment benchmark, we reassigned stated profiles across participants 1,000 times to test whether the correct person--profile link yields a larger mean coefficient than random links.

To compare representations, we estimated the moderation for each representation separately and for all three jointly, with the main effects of lens composites and profiles and model fixed effects in every model. Joint coefficients are conditional on the other representations; separate coefficients show what each representation captures on its own. To carry the uncertainty of constructing the representations into the inference, we also bootstrapped participants through the whole construction and estimation (200 draws), repeated the analysis over 50 random half-splits, and applied Holm correction to the three pairwise contrasts.

\textbf{Concrete feedback.} The length weight was allowed to vary with a participant's net share of conversations containing requests for shorter rather than longer responses, computed over their other conversations and, prospectively, over the conversations that preceded the rated one. The four main tests in this exploratory analysis (same conversation, other conversations, earlier conversations, and formatting mentions for the weight on lists and headings) were Holm-corrected together. A common model entered the concrete-feedback interaction and the seven priority interactions for the length weight together, so that the two kinds of self-report could be compared directly.

\textbf{Reasons.} For each conversation and attribute, the reported reason was regressed on chosen-minus-other feature differences (M1), adding perceived performance on the same attribute (M2) and then mean perceived performance on the other attributes (M3). All variables were demeaned within participant and attribute. A separate model entered perceived performance on all seven attributes to assess specificity; its sample was restricted to complete cases before demeaning.

\textbf{Individual differences.} For each feature, we fitted a linear mixed model to orthonormal within-conversation contrasts of the ratings, which remove the conversation means while keeping the residual covariance valid. The model has a participant-level and a conversation-level random slope on that feature, with all features and model and position effects as fixed effects. It was fitted by restricted maximum likelihood (REML), re-estimating the fixed effects by generalized least squares at every value of the variance components.

The participant component was tested against the conversation-only model with a likelihood ratio referred to the 50:50 mixture of $\chi^2_0$ and $\chi^2_1$, with Holm correction across the 12 features. We checked the calibration of this reference distribution by parametric bootstrap under the fitted null for four features, under person-specific residual variances for length, and under the fitted alternative for length. The primary scale divides each rating by the participant's standard deviation of within-conversation rating deviations, which adjusts for differences in rating dispersion; raw ratings are reported alongside.

We also estimated the variance of person-stable slopes from the covariance of each participant's slopes in two random halves of their conversations, averaged over 50 random splits, with a participant bootstrap that recomputes the same 50-split average. Finally, we simulated the split-half procedure on the actual design, setting each feature's participant-level standard deviation to zero, its fitted value or twice that value while holding the conversation-level standard deviation at its fitted value (100 simulated datasets per scenario, each summarized over 10 random splits).

\textbf{Prediction.} For participants with at least four conversations (\NpredEligible{}; \NpredPersons{} have response pairs with different ratings), we predicted the ordering of first-turn responses in each held-out conversation. All information sources were evaluated using the same ten participant folds, repeated three times. The main baseline was a pooled linear model of ratings with all response features except similarity to the participant's own texts, similarity to a common reference pool of training participants' texts, and model and position effects. Each information source was added to this baseline; a supplementary nonlinear comparison added stated priorities to a pooled gradient-boosting baseline.

All quantities learned across participants (feature scaling and winsorizing, pooled weights, weights linking person-level information to features, aggregate profiles, the reference pool, and every shrinkage choice) were estimated from the outer-training participants only, with shrinkage chosen on an inner split of those participants. Person-specific weights were learned from the participant's other conversations and shrunk toward the pooled weights; a forward-only version used only earlier conversations.

Pairwise gradient boosting was trained on both orientations of every pair with complementary labels and scored with a symmetric rule, and a linear pairwise logistic model with the same inputs separates the pairwise objective from the added flexibility.

The primary metric is pairwise accuracy, the share of response pairs with different ratings that are ordered correctly, computed for each participant and averaged across participants. Top-choice accuracy and squared rating error are secondary. Confidence intervals were obtained by bootstrapping participants while holding the fitted predictions fixed.

\subsection{Analysis plan and transparency}
\label{sec:plan}

Before estimating any outcome model, we wrote a time-stamped pre-analysis plan that specified the sample, feature definitions, primary tests and robustness checks; the plan was not publicly registered. Before it was written, feasibility checks had computed descriptive statistics, correlations between stated priorities and reasons and between reasons and perceived performance (largely reproducing \citealp{kirk2024prism}), and one pooled association between rating and response length; length therefore enters as a pre-specified feature rather than a discovery. H1, H2, H4 and the prediction contest are planned analyses. The aggregate-profile analysis was planned as a secondary check. The comparison of representations, the perceptual-correspondence analysis, the lens composites, the coding of concrete feedback, the individual-difference models, the context interactions and the full prediction protocol were added after the plan and are exploratory. Deviations and additions are listed in Appendix~\ref{app:deviations}.

\section{Results}

The results address two linked questions: what preference reports reveal about evaluation, and how much that information helps predict individual choices. We begin with the features associated with ratings and perceived attributes, then examine stated priorities, concrete feedback and reasons. The final analyses distinguish stable differences in evaluation from the ability to recover and use them.

\subsection{Features associated with ratings}
\label{sec:assoc}

\begin{figure}[t]
\centering
\includegraphics[width=\textwidth]{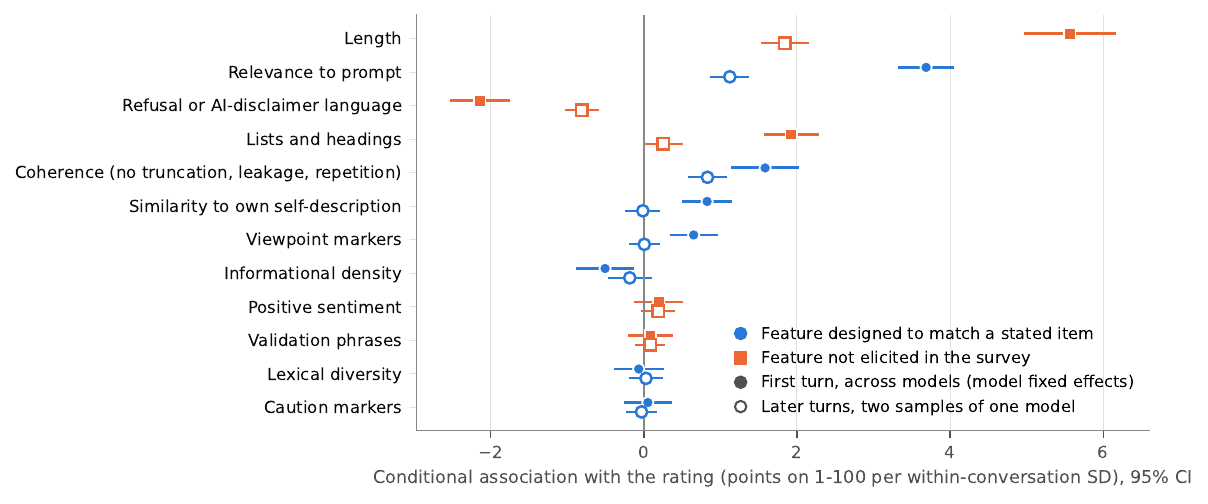}
\caption{Conditional associations of response features with ratings. Filled markers: first turn (Equation~\ref{eq:t1}). Hollow markers: later turns, within pairs of responses sampled from the same model. Bars are 95\% confidence intervals clustered by participant.}
\label{fig:assoc}
\end{figure}

Conditional on prompt, producing model and position, the strongest association was with a feature that the survey did not elicit, response length (Figure~\ref{fig:assoc}). A one-standard-deviation increase in length was associated with \TOneLen{} more rating points (SE \TOneLenSE{}), lists and headings with \TOneStruct{} points (SE \TOneStructSE{}), and refusal or disclaimer language with \TOneRef{} points (SE \TOneRefSE{}). Among the text proxies, relevance to the prompt (\TOneHelp{}, SE \TOneHelpSE{}) and coherence (\TOneFlu{}, SE \TOneFluSE{}) carried substantial weight, and caution markers carried almost none (\TOneSafe{}, SE \TOneSafeSE{}). The non-elicited features were jointly significant ($\chi^2(5) = \TOneWald{}$) and added at least as much explained within-conversation variance as the seven text proxies. The within-conversation $R^2$ was \RtwoFE{} with model and position effects only, \RtwoMatched{} after adding the text proxies, \RtwoUnclaimed{} after adding the non-elicited features and \RtwoAll{} with both. These associations support H2.

In later turns, where participants rated two samples from the same model, the first-turn weights predicted which sample a participant preferred in \WMAcc\% of \WMPairs{} pairs (participant-bootstrap SE \WMAccSE{} percentage points, chance 50\%), and length (\WMLen{}, SE \WMLenSE{}), relevance (\WMHelp{}), coherence (\WMFlu{}) and refusal language (\WMRef{}) kept their direction. Similarity to the self-description (\WMVal{}) and viewpoint markers (\WMDiv{}) did not. Feature weights also varied with the request in ways shared across participants (Table~\ref{tab:context}). The length weight rose by \CtxLenPrompt{} points per standard deviation of prompt length and by \CtxLenContro{} in conversations about controversial topics; the penalty for refusal language was smaller in those conversations (\CtxRefContro{}), and the weight on relevance to the prompt was lower (\CtxHelpContro{}).

\subsection{Concrete features are associated with perceived attributes}
\label{sec:cv}

\begin{figure}[t]
\centering
\includegraphics[width=\textwidth]{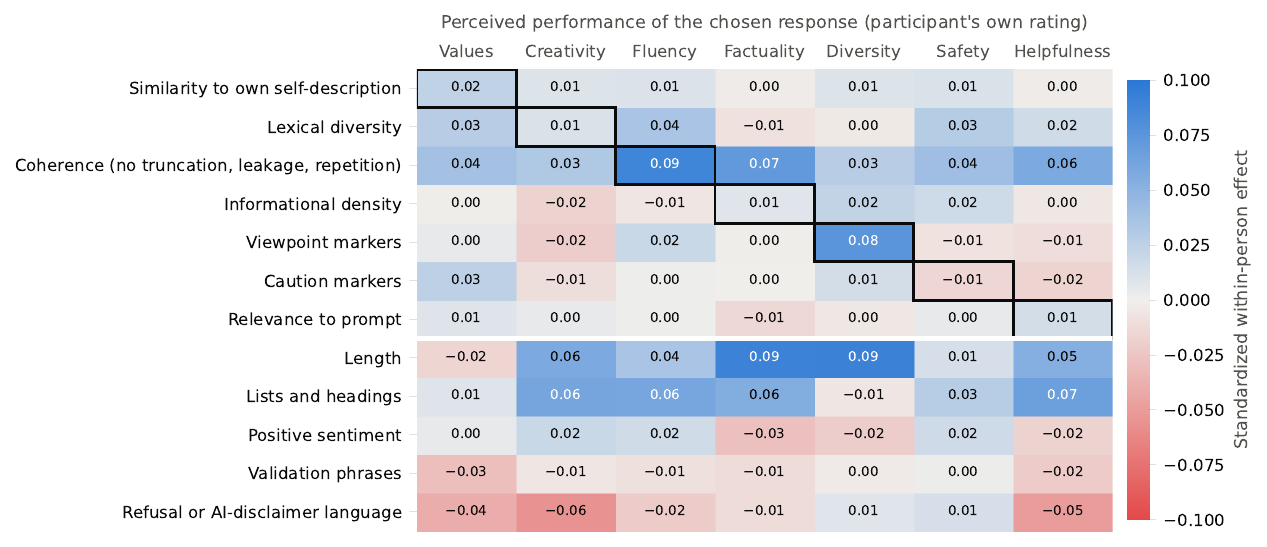}
\caption{How response features relate to perceived attributes. Cells are standardized partial within-participant associations of the chosen-minus-other feature difference with the participant's perceived performance of the chosen response (one regression per attribute with all features; $n \ge \CVN$ conversations each). Outlined cells pair each designed feature with the attribute it was meant to approximate. Features below the white line were not elicited in the survey.}
\label{fig:cv}
\end{figure}

The prominence of length, formatting and refusal language raises a measurement question: how do these features relate to the attributes participants say they perceive? Figure~\ref{fig:cv} relates each feature to participants' own perceived performance of the chosen response, within participant. Two text proxies showed a clear partial association with the attribute they were meant to approximate: coherence with perceived coherence ($\beta = \CVFlu$) and viewpoint markers with perceived viewpoint diversity ($\beta = \CVDiv$). Informational density (\CVFactFact{} for perceived factuality), relevance to the prompt (\CVHelpHelp{} for perceived helpfulness) and caution markers (\CVSafeSafe{} for perceived safety) showed little partial association with their attributes. The features the survey did not elicit were instead among the strongest correlates of perceived attributes. Longer responses were perceived as more factual and informative (\CVLenFact{}) and as summarizing more viewpoints (\CVLenDiv{}); lists and headings were associated with higher perceived helpfulness (\CVStructHelp{}) and creativity (\CVStructCre{}); and refusal language with lower perceived creativity (\CVRefCre{}) and helpfulness (\CVRefHelp{}).

The pattern did not depend on using relative contrasts (Table~\ref{tab:cvsens}). With the chosen response's own feature levels, controlling for conversation type and prompt length, the strongest correlate was a non-elicited feature for \CVNTopNonElicitedLv{} of the 7 attributes (\CVNTopNonElicited{} with contrasts). Length remained the strongest correlate of perceived factuality (\CVLvLenFact{}) and diversity (\CVLvLenDiv{}), and refusal language of perceived helpfulness (\CVLvRefHelp{}) and creativity (\CVLvRefCre{}). Relevance to the prompt (\CVLvHelpHelp{}) and similarity to the self-description (\CVLvValVal{}) corresponded somewhat more closely to their attributes in this version. Perceived attributes thus covary with concrete features that the attribute names do not mention, and most text proxies correspond only weakly to the perceptions they were designed to approximate.

\subsection{Stated priorities and the weights on text proxies}
\label{sec:h1}

The planned test of alignment (H1) asks whether participants who rate an attribute as more important give more weight to the text proxy designed for it. The mean moderation coefficient across the seven text proxies was small and not statistically distinguishable from zero (\TTwoGamma{}, SE \TTwoSE{}, $p = \TTwoP$; 95\% CI \TTwoCIlo{} to \TTwoCIhi{}; Table~\ref{tab:robust}), and the joint test of the seven coefficients was not significant ($\chi^2(7) = \TTwoWald$, $p = \TTwoWaldP$). At the attribute level, the most consistent signal concerned helpfulness and its proxy, relevance to the prompt: participants who rated helpfulness as more important weighted relevance more (ipsatized \IpsHelp{}, $p = \IpsHelpP$, $p_{\mathrm{Holm}} = \IpsHelpHolm$; raw importance \RawHelp{}, SE \RawHelpSE{}, $p_{\mathrm{Holm}} = \RawHelpHolm$; Table~\ref{tab:attlevel}).

One alternative specification, which replaces coherence and similarity to the self-description with readability and similarity to the participant's free-text wishes, gave a significant joint test ($p = \AltWaldP$) around a mean near zero, so some individual coefficients in that specification differ from zero even though their average does not. In the profile-reassignment benchmark, the correct person--profile links did not produce a reliably larger mean coefficient than random links ($p = \TTwoPermP$). The mean coefficient stayed between \RobMin{} and \RobMax{} across specifications, and between \LomoMin{} and \LomoMax{} when each model was left out with the conversation transformations recomputed. H1 is not supported for the mean of the text proxies.

\begin{table}[t]
\centering
\caption{Mean moderation of the weights on the designed text proxies by stated importance (ipsatized unless stated), across specifications. Joint $p$: seven-degree-of-freedom Wald test of all moderation coefficients. SEs clustered by participant.}
\label{tab:robust}
\small
\begin{tabular}{>{\raggedright\arraybackslash}p{8.2cm}rrrr}
\toprule
Specification & Mean & SE & $p$ & Joint $p$ \\
\midrule
Primary (ipsatized, all analysed responses) & 0.09 & 0.08 & 0.304 & 0.151 \\
Four-response conversations only & 0.09 & 0.09 & 0.311 & 0.407 \\
PRISM balanced subset & 0.05 & 0.09 & 0.567 & 0.424 \\
US census-representative sample & 0.07 & 0.22 & 0.740 & 0.767 \\
UK census-representative sample & 0.08 & 0.19 & 0.665 & 0.926 \\
Excluding the 10\% of participants with the largest same-model score gaps & 0.14 & 0.09 & 0.102 & 0.057 \\
Within-person standardized ratings (SD units) & 0.004 & 0.003 & 0.224 & 0.375 \\
Without model fixed effects (position effects kept) & 0.02 & 0.09 & 0.800 & 0.289 \\
Alternative features (readability; similarity to own wishes), with their main effects & $-$0.04 & 0.09 & 0.657 & 0.021 \\
Leave one model out, re-demeaned (21 runs): range of the mean & [0.02, 0.11] & & & \\
\bottomrule
\end{tabular}
\end{table}

\subsection{Alignment depends on how attributes are operationalized}
\label{sec:levels}

\begin{figure}[t]
\centering
\includegraphics[width=\textwidth]{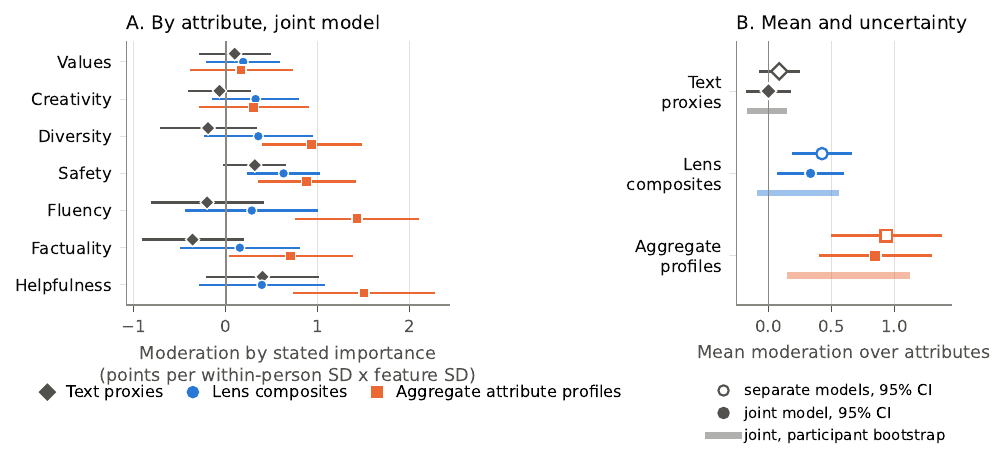}
\caption{Alignment between stated priorities and ratings under three representations of the same attributes. (A) Moderation of each attribute's weight by stated importance (ipsatized), joint model, 95\% confidence intervals; grey diamonds: text proxies; blue circles: lens composites; orange squares: aggregate attribute profiles. (B) Mean over attributes: hollow markers, separate models; filled markers, joint model (95\% confidence intervals); shaded bars, 95\% intervals from a participant bootstrap through the construction of the representations. Text proxies: the designed features of Table~\ref{tab:cues}. Lens composites: features weighted by how they predict perceived performance, estimated on the other half of participants. Aggregate attribute profiles: other participants' perceived performance of the producing model, from the other half and corrected for raters' stated priorities.}
\label{fig:levels}
\end{figure}

The weak perceptual correspondence of most text proxies makes measurement a plausible explanation for the small average in H1. We therefore compared the three representations of the same attributes. The moderation of rating weights by stated priorities depended on how the attributes were represented (Figure~\ref{fig:levels}; Table~\ref{tab:levels}). In the joint model, the mean moderation coefficient was \LevProx{} for text proxies (SE \LevProxSE{}, \LevProxP), \LevLens{} for lens composites (SE \LevLensSE{}, \LevLensP) and \LevProf{} for aggregate attribute profiles (SE \LevProfSE{}, \LevProfP). Estimated separately, the three means were \LevSepProx{} (\LevSepProxP), \LevSepLens{} (\LevSepLensP) and \LevSepProf{} (\LevSepProfP). Profiles exceeded proxies by \DiffProfProx{} (SE \DiffProfProxSE{}, \DiffProfProxHolm{} across the three contrasts; Table~\ref{tab:levcontr}). The lens composites lay between the two, and neither of their contrasts was distinguishable from zero after correction (profiles minus lens \DiffProfLens{}, \DiffProfLensHolm{}; lens minus proxies \DiffLensProx{}, \DiffLensProxHolm{}).

The profile--proxy contrast was consistent across resampling checks. Profiles exceeded proxies in \RepShareProfGtProx\% of \RepN{} random half-splits (smallest difference \RepDiffMin{}), and a participant bootstrap that repeated the construction of lens weights, echo corrections and profiles gave a 95\% interval of \DiffProfProxBootLo{} to \DiffProfProxBootHi{} for the contrast. The same bootstrap gave \LevBootProfLo{} to \LevBootProfHi{} for profiles and \LevBootLensLo{} to \LevBootLensHi{} for lens composites.

Two checks address whether the profile result reflects a general quality dimension rather than specific attributes. The seven profiles are strongly correlated across models, and their mean, which we call common perceived quality, is reproducible across halves of raters ($r = \QualRel$). After common perceived quality was removed, the attribute-specific deviations still moderated the weights (mean \SpecDev{}, SE \SpecDevSE{}, \SpecDevP). Pairing each priority with the profile of a different attribute gave a mean coefficient of \SpecMisMean{} over six cyclic mismatches (largest \SpecMisMax{}), against \SpecMatched{} for the correct pairing (Table~\ref{tab:profspec}). Participants who rated an attribute as relatively important thus gave relatively higher ratings to models that other participants perceived as strong on that specific attribute.

Profiles were moderately to highly reproducible across halves of raters (split-half $r$ from \ProfRelLo{} to \ProfRelHi{}), whereas lens composites predicted held-out perceptions weakly (within-person $r$ from \LensHeldLo{} to \LensHeldHi{}; Table~\ref{tab:levrel}). The profile result also held across the construction choices in Table~\ref{tab:profiles}, including a disjoint-half version with model fixed effects (\ConsDis{}, SE \ConsDisSE{}).

\subsection{Concrete feedback tracks the length weight}
\label{sec:concrete}

The profile result concerns attributes that the survey named. The features that carried the most weight, above all length, were never named in the survey, but participants raised them in their own words. The coding rules detected at least one request for shorter or more concise responses from \ConcShortPart\% of participants and at least one request for longer or more detailed ones from \ConcLongPart\%. They detected praise of brevity from \ConcBrevPraisePart\% of participants, praise of detail from \ConcDetPraisePart\%, mentions of formatting from \ConcFmtPart\% and complaints about refusal or evasion from \ConcRefPart\%.

Requests about length tracked the weight attached to length (Table~\ref{tab:concrete}). Participants whose other conversations contained more requests for shorter rather than longer responses gave length less weight. A one-standard-deviation increase in the net share of such conversations changed the length weight by \ConcLenTSD{} rating points per standard deviation of length. On the original net-share scale, the coefficient was \ConcLenT{} (SE \ConcLenTSE{}, \ConcLenTHolm{} across the four main exploratory tests). The association held prospectively, using only the conversations that preceded the rated one (\ConcLenE{}, SE \ConcLenESE{}, \ConcLenEHolm{}), and when praise of brevity or detail was added to the score (Table~\ref{tab:concrete}). Requests made in the same conversation, written after its ratings, showed no detectable association with its length weight (\ConcLenC{}, \ConcLenCP). The estimated association of formatting mentions with the weight on lists and headings was positive but uncertain (\ConcFmtT{}, $p_{\mathrm{Holm}} = \ConcFmtTHolm$). Refusal complaints were too few for an interaction analysis and are reported descriptively.

We compared this concrete-feedback signal with the information supplied by stated priorities. In a common model with both blocks, the concrete-feedback interaction remained similar (\CommonIpsConc{}, SE \CommonIpsConcSE{}, \CommonIpsConcP), whereas the relative priorities, entered as a full-rank basis of \CommonIpsPriDf{} columns because the seven ipsatized scores sum to zero, added no detectable information ($p = \CommonIpsPriP$; Table~\ref{tab:common}). The absolute-importance interactions did contribute ($p = \CommonRawPriP$), mainly because participants who rated values alignment as more important rewarded length less (\RawValLen{}, \RawValLenP), an unanticipated pattern. On their own, relative priorities showed no detectable moderation of the weights on length, refusal language or lists and headings (smallest $p = \RefAttrIpsP$; Table~\ref{tab:attrconcrete}), and none of a small set of theory-motivated pairings of priorities with non-elicited features survived correction (smallest $p_{\mathrm{Holm}} = \CrossMinHolm$; Table~\ref{tab:crosslevel}). Requests about length thus carried information about the weight attached to length beyond the relative profile of stated priorities.

\subsection{Reasons track perceived attributes}
\label{sec:reasons}

The preceding analyses concern reports across conversations. Reasons instead describe a particular choice. These reports were weakly associated with the measured feature differences (Table~\ref{tab:reasons}). With feature differences alone (M1), the standardized association was \RsnMoneCue{} for the matched feature and \RsnMoneLen{} for length. Adding perceived performance on the same attribute (M2, $\beta = \RsnMtwoOwn$) reduced the feature associations to \RsnMtwoCue{} and \RsnMtwoLen{}, and perceived performance on the other attributes added a smaller association (M3, $\beta = \RsnMthreeOthers$). The reasons were attribute-specific: in a model with all seven perceptions, the same-attribute association ranged from \SpecOwnLo{} to \SpecOwnHi{}, whereas the mean cross-attribute association ranged from \SpecCrossLo{} to \SpecCrossHi{} (Table~\ref{tab:specificity}). Reasons were therefore more strongly associated with perceived performance on the named attribute than with the measured feature differences (H4).

\begin{table}[t]
\centering
\caption{Standardized associations with the reported reason. Conversation $\times$ attribute observations; variables demeaned within participant and attribute. $^{*}$ $p < 0.05$ (participant-clustered).}
\label{tab:reasons}
\small
\begin{tabular}{lrrr}
\toprule
Standardized association with the reported reason & M1 & M2 & M3 \\
\midrule
Text-proxy contrast (chosen minus others) & 0.031$^{*}$ & 0.013$^{*}$ & 0.013$^{*}$ \\
Length contrast & 0.045$^{*}$ & 0.026$^{*}$ & 0.023$^{*}$ \\
Lists and headings contrast & 0.021$^{*}$ & 0.003 & $-$0.001 \\
Positive sentiment contrast & 0.005 & 0.005 & 0.005 \\
Validation phrases contrast & $-$0.015 & $-$0.008 & $-$0.006 \\
Refusal language contrast & $-$0.013 & $-$0.003 & $-$0.001 \\
Perceived performance, same attribute &  & 0.547$^{*}$ & 0.478$^{*}$ \\
Perceived performance, other attributes (mean) &  &  & 0.121$^{*}$ \\
\bottomrule
\end{tabular}
\end{table}

\subsection{Stable individual differences and their recoverability}
\label{sec:indiv}

\begin{figure}[t]
\centering
\includegraphics[width=\textwidth]{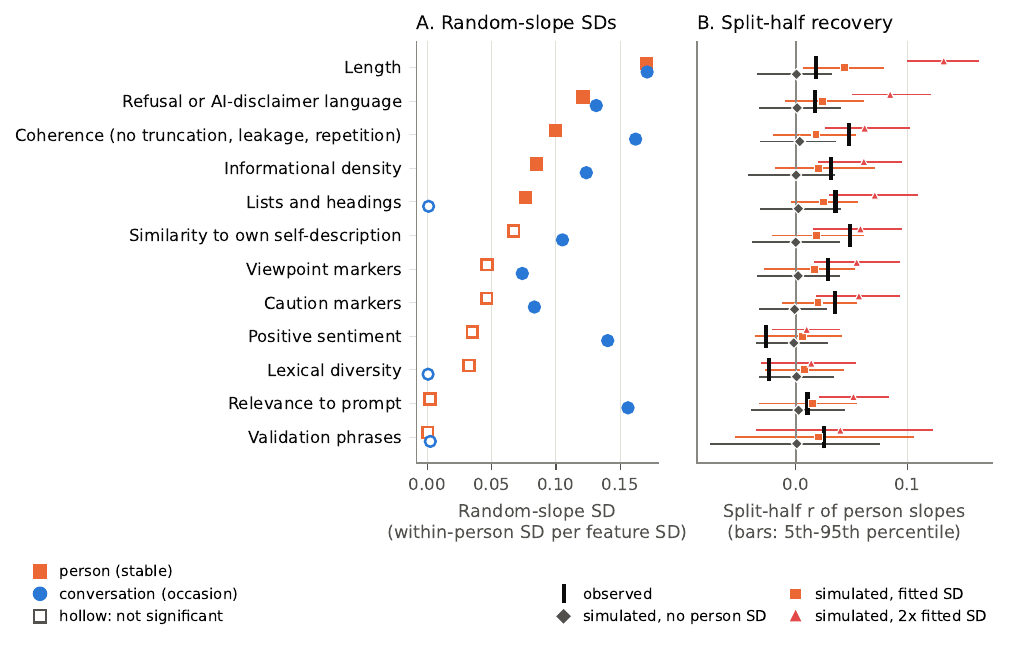}
\caption{Individual differences. (A) Standard deviations of participant-level (stable) and conversation-level (occasion) random slopes, estimated jointly by REML on orthonormal within-conversation contrasts of within-person standardized ratings; hollow markers are not significant (participant level: Holm-corrected across features). (B) Split-half correlation of participant-level slopes, observed (black ticks) and simulated on the actual design with participant-level standard deviations set to zero, to their fitted values, or to twice those values (conversation-level standard deviations at their fitted values; 100 simulated datasets per scenario; markers are means and bars span the 5th to 95th percentiles of simulated datasets).}
\label{fig:indiv}
\end{figure}

The moderation analyses show which reports track differences in evaluation. We next ask how much stable variation in feature weights exists, and how precisely it can be recovered for each participant. Participants differed stably in how they weighted several concrete features (Figure~\ref{fig:indiv}A; Table~\ref{tab:likelihood}). In the REML model with participant-level and conversation-level random slopes, the participant-level component was significant for \NSigPerson{} of the 12 features without adjustment and for \NSigPersonHolm{} after Holm correction. It was largest for length, with a participant-level standard deviation of \TauPLen{} (in within-person rating standard deviations per feature standard deviation; likelihood ratio \LRLen{}) against a pooled weight of \BetaStdLen{} on the same scale. Refusal language (\TauPRef{}), coherence (\TauPFlu{}), informational density (\TauPFact{}) and lists and headings (\TauPStruct{}) followed. Conversation-level components were as large or larger for most features, consistent with the context dependence in Section~\ref{sec:assoc}.

Simulation checks supported the participant-level signal under the fitted model assumptions (Table~\ref{tab:remlcal}). In \CalNSim{} datasets simulated under the fitted null for each of four features, the 5\% test rejected in \CalRejLo\% to \CalRejHi\% of datasets, and the largest simulated likelihood ratio, \CalLRNullMax{}, was below each of the four observed statistics (smallest \CalLRObsMin{}; length \LRLen{}, refusal language \LRRef{}). Allowing the residual variance to differ between participants raised the null rejection rate for length to \CalRejHet\%, while the largest simulated statistic (\CalLRHetMax{}) remained far below the observed one. With the participant-level standard deviation for length set to its fitted value (\CalTauTrue{}), the test rejected the null in \CalPowerCount{} of \CalNNon{} simulated datasets and recovered the size of the component (mean estimate \CalTauHat{}).

A separate check examined whether stable differences reflected recurring differences in participants' prompts. Because the length weight rises with prompt length, someone who habitually writes long prompts could also show a consistently stronger length weight. Adding the context interactions left the participant-level component for length almost unchanged (\TauPLenCtx{}, likelihood ratio \LRLenCtx{}), and \NSigPersonCtx{} features retained a significant component after Holm correction. A moment estimator that relies on fewer distributional assumptions is much less precise: its 95\% interval for the person-stable variance of the length weight ran from \MomLenLo{} to \MomLenHi{} squared rating points, and it excluded zero for \MomNExcl{} of the 12 features (Table~\ref{tab:moment}). The evidence for stable differences therefore rests on the calibrated likelihood analysis.

Despite this population-level evidence of stable variation, estimates for individual participants were poorly reproducible across their conversations. Slopes estimated in two random halves of a participant's conversations correlated at \SHrMean{} on average (range across features \SHrLo{} to \SHrHi{}). The recovery simulation reproduced this low reliability at the fitted levels of heterogeneity (Figure~\ref{fig:indiv}B; Table~\ref{tab:moment}). With every feature's participant-level and conversation-level standard deviations set to their fitted values, the simulated split-half correlations ranged from \SimFitLo{} to \SimFitHi{} (\SimFitLen{} for length). Doubling the participant-level standard deviations raised them to at most \SimTwoHi{}, and without participant-level variation they stayed within \SimNullAbsMax{} of zero (\SimNRep{} simulated datasets per scenario; Monte Carlo SE at most \SimMCSEMax{}). Low split-half reproducibility is therefore expected in this design even when stable differences of the estimated size exist. Reported reasons were far more reproducible across the same random splits, with uncorrected split-half correlations of \RelReasonRLo{} to \RelReasonRHi{} (\RelReasonIpsRLo{} to \RelReasonIpsRHi{} for the within-person profile), indicating that reproducibility differed substantially between repeated reports and estimated feature weights.

\subsection{Predicting held-out rankings}
\label{sec:predict}

\begin{figure}[t]
\centering
\includegraphics[width=\textwidth]{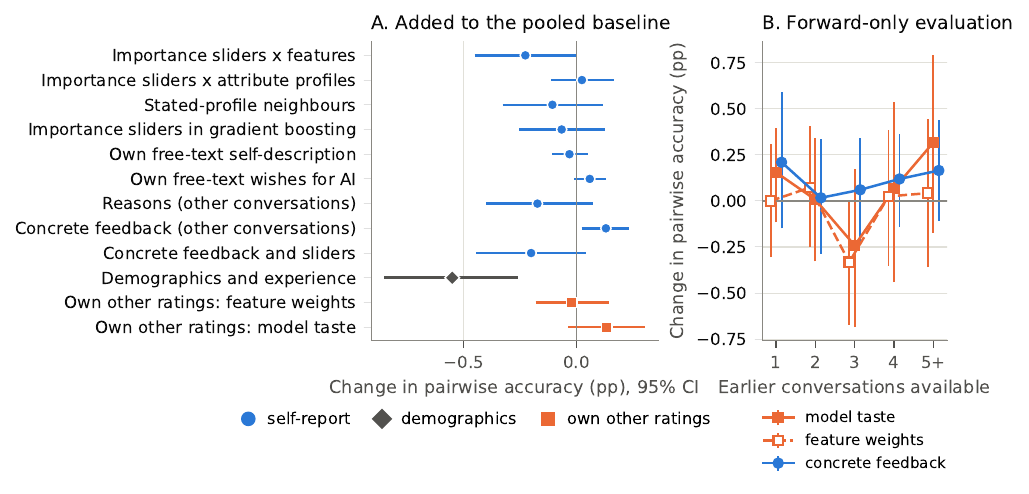}
\caption{Prediction of held-out first-turn rankings. (A) Change in participant-averaged pairwise accuracy when each information source is added to the pooled linear baseline (\BaseAcc\%); the boosting-based priorities comparison uses a pooled boosting baseline. Bars are 95\% participant-bootstrap intervals (three repetitions of ten participant folds). Blue: self-reported information; grey: demographics and experience; orange: the participant's own other ratings. (B) Forward-only evaluation: change in accuracy when person-specific information comes only from earlier conversations, by number of earlier conversations (reasons are in Table~\ref{tab:chrono}).}
\label{fig:predict}
\end{figure}

Poor recovery of individual feature weights raises a practical question: can either self-reports or observed ratings improve prediction over a pooled model? The pooled baseline ordered held-out responses correctly in \BaseAcc\% of pairs (Figure~\ref{fig:predict}; Table~\ref{tab:predict}), and none of the four ways of incorporating stated priorities produced a detectable improvement over its corresponding pooled baseline. Relative to the linear baseline, sliders interacting with features changed accuracy by \GSlider{} percentage points (95\% CI \GSliderLo{} to \GSliderHi{}), sliders interacting with aggregate profiles by \GProf{} (\GProfLo{} to \GProfHi{}), and a model trained on the participants with the most similar stated profiles by \GNbr{}. Adding sliders to the pooled gradient-boosting model, which can represent nonlinear interactions, changed accuracy by \GGBM{} relative to that baseline. Reasons from other conversations (\GReason{}) and person-specific feature weights learned from the participant's other conversations (\GOwnCue{}) produced no detectable improvement in ranking accuracy, and demographics with experience measures lowered it (\GDemo{}, \GDemoLo{} to \GDemoHi{}).

Three sources gave the largest point estimates: concrete feedback from other conversations (\GConc{}, \GConcLo{} to \GConcHi{}), the only addition whose interval lay above zero; person-specific taste for models (\GOwnModel{}, \GOwnModelLo{} to \GOwnModelHi{}); and similarity to participants' own free-text wishes for an AI (\GFTWish{}, \GFTWishLo{} to \GFTWishHi{}). These intervals are not corrected for the \NAdditions{} comparisons, and paired comparisons did not separate the wishes from concrete feedback or model taste (Table~\ref{tab:watchask}).

Direct comparisons showed that concrete feedback outperformed sliders interacting with features (\PDConcSlid{}, \PDConcSlidLo{} to \PDConcSlidHi{}) but not sliders interacting with profiles (\PDConcProf{}, \PDConcProfLo{} to \PDConcProfHi{}), and adding the feature-interacting sliders to concrete feedback lowered accuracy (\PDBothConc{}, \PDBothConcLo{} to \PDBothConcHi{}). The largest upper endpoint among the individual 95\% intervals was \GHiMax{} percentage points. The direct comparison of watching with asking (H5) gave limited support: the interval excluded zero in \WANSig{} of \WANTot{} comparisons, person-specific model taste against sliders interacting with features (\WAMax{}, \WAMaxLo{} to \WAMaxHi{}), mainly because those sliders lowered accuracy.

Squared rating error gave a different picture. Person-specific feature weights (\GOwnCueMSE{} squared rating points, \GOwnCueMSELo{} to \GOwnCueMSEHi{}) and sliders interacting with profiles (\GProfMSE{}, \GProfMSELo{} to \GProfMSEHi{}) reduced it, although neither changed the ordering of responses detectably, so individual information improved the prediction of rating levels more than of rankings.

Improving the pooled prediction pipeline produced a larger ranking gain than adding personal information. Replacing the linear pooled model with a pooled symmetric gradient-boosting model, without any personal information, raised pairwise accuracy by \GBMPool{} points (\GBMPoolLo{} to \GBMPoolHi{}). A comparison holding the linear functional form constant also favoured a pairwise objective: a linear pairwise logistic model with the same inputs gained \LinPairPool{} (\LinPairPoolLo{} to \LinPairPoolHi{}), and boosting exceeded it by \GBMvsLin{} (\GBMvsLinLo{} to \GBMvsLinHi{}).

The forward-only evaluation gave a similar result: using only earlier conversations produced small changes whose intervals mostly included zero, and no clear increase with the number of earlier conversations was observed (Table~\ref{tab:chrono}; estimated trend for model taste \TrModel{} points per earlier conversation, \TrModelLo{} to \TrModelHi{}); with five or more earlier conversations, person-specific model taste changed accuracy by \ChModelFive{} (\ChModelFiveLo{} to \ChModelFiveHi{}).

\subsection{Experience and item order}

Two further checks concern the elicitation itself. We found no detectable difference in alignment by familiarity with language models or frequency of use: direct contrasts between experience groups were not significant (smallest $p = \ExpMinP$; Table~\ref{tab:experience}). The estimated shifts in stated importance across randomized item positions were at most \OrderMax{} points, with no significant joint effect (Table~\ref{tab:order}), providing no clear evidence of an item-order effect.

\section{Discussion}

\subsection{The explanations revisited}

The central finding is that the correspondence between stated priorities and response ratings depends on how the named attributes are measured. Mean alignment was near zero for designed text proxies and positive for aggregate attribute profiles derived from other participants' perceptions. The profile--proxy contrast survived correction for multiple comparisons and resampling through the construction of the representations. Lens composites lay between these endpoints, with neither pairwise contrast statistically resolved. These results make the interpretation of a stated--revealed preference gap contingent on the measure used to represent the stated attribute.

The competing explanations in Table~\ref{tab:explanations} help interpret this pattern. Measurement mismatch is plausible because most text proxies corresponded weakly to the perceptions they were intended to approximate, while length, formatting and refusal language covaried with those perceptions. The profile result also contained attribute-specific information: alignment persisted after removing common perceived quality, and correctly matched priority--profile pairs had larger mean coefficients than the mismatched pairings. However, profiles aggregate many raters and chosen responses, whereas lens composites predicted held-out perceptions weakly. Reliability, aggregation and selection therefore remain part of the explanation. The evidence supports sensitivity to operationalization; level of description provides a theoretical account to test directly. Context dependence contributes as well, since feature weights varied systematically with the request.

The comparison of self-reports makes this interpretation more concrete. Participants' requests for shorter or longer answers tracked the weight they attached to length, including when the requests came from earlier conversations. This association remained after accounting for relative priorities across the seven abstract attributes. Choice-specific reasons, in turn, tracked perceived performance on the named attributes more closely than the measured feature differences. Each report thus provided a different view of evaluation: priorities related to aggregate perceived attributes, requests about length to the length weight, and reasons to perceptions accompanying a particular choice.

\subsection{Theoretical implications}

These findings apply the lens model of judgment \citep{hammond1955probabilistic,karelaia2008determinants} to the interpretation of preference reports in human--AI evaluation. Observable response features, perceived attributes and stated priorities are related but distinct quantities. A weak association between priorities and a designed feature can therefore coexist with a stronger association between those priorities and a perception-based measure. This helps interpret the weak stated--revealed correspondence reported in PRISM \citep{kirk2024prism} and in creative writing \citep{chung2025literarytaste}: the size of the gap depends partly on what is compared.

The findings also sharpen the question of self-insight \citep{nisbett1977telling}. Reasons were informative about reported perceptions, while conveying less about the measured features of a response. This is compatible with a distinction between abstract and concrete representations \citep{trope2010construal}, and motivates tests that hold content and measurement quality constant while varying the level of elicitation. A separate contribution concerns individual differences. Stable variation in evaluation existed at the population level, yet short histories yielded poorly reproducible estimates of individual feature weights. Establishing heterogeneity and estimating a useful personal profile are distinct empirical tasks.

\subsection{Practical implications}

The findings suggest three priorities for preference elicitation and personalization. First, the form of elicitation should match the decision it is intended to inform. Relative priorities across abstract attributes aligned with aggregate attribute profiles but revealed little about the weight attached to length. Concrete requests supplied additional information about that weight. Direct questions about response length and other observable properties are therefore promising candidates for evaluation against held-out choices.

Second, because concrete features such as length covary with perceived attributes, ratings collected under attribute labels can reward properties that the labels do not mention \citep{singhal2023long,dubois2024length}. Attribute-level reasons collected with ratings tracked perceptions closely and features weakly, so an audit of the concrete properties associated with ratings requires direct analysis of those properties alongside the reasons.

Third, with a median of \MedConv{} conversations per participant, the largest estimated improvement in held-out pairwise accuracy from personal information was \GMax{} percentage points, even though stable individual differences existed, whereas a stronger pooled prediction pipeline added \GBMPool{} points. Person-specific weights also improved the prediction of rating levels without improving rankings. Demographic information was no substitute either: it lowered held-out accuracy here, and LLMs prompted with demographic profiles fail to beat non-LLM baselines at predicting individual survey responses \citep{chen2026synthetic}. Personalization methods should therefore be evaluated against a strong pooled baseline, on held-out choices from the same person and with metrics that match their purpose, and the amount of history needed to estimate individual weights should be treated as a design parameter.

\subsection{Limitations and future research}

The main measurement constraint is the use of transparent text proxies without independent annotation of the intended attributes. Perceptual correspondence describes how these proxies relate to participants' post-conversation ratings of chosen responses, which cluster near the ceiling. Independent annotation, blind to ratings and stated priorities, would establish how well the proxies capture factual correctness and the appropriateness of caution or refusal. The independent validation of the concrete-feedback codes found missed requests, especially indirect requests for more detail, as well as some false positives, and it relied on a single coder. The validation quantifies the uncertainty in these codes but does not determine the direction of bias in their associations with ratings: the net score for length requests combines two codes with different error rates, and even when misclassification biases estimates toward zero on average, a particular estimate can lie farther from zero than the true association \citep{jurek2005proper}. Broader rules for indirect requests, validated by several independent coders on a larger sample, would refine the prevalence estimates and support a measurement-error model for the concrete-feedback associations.

The observational design supports conditional associations among features, perceptions and ratings. Perceptions and reasons were collected together after the conversation and referred to the chosen response. Their close correspondence can reflect shared perceptions, shared wording or post-choice reconstruction. Aggregate profiles also summarize chosen responses and are strongly correlated across the 21 models (mean correlation \ConsCorr{}). Sample splitting separates profile construction from evaluation, but aggregation and selection remain relevant to interpreting the profile advantage. Concrete feedback likewise reflects the responses and tasks participants encountered; its earlier-conversation association establishes temporal ordering without equating it to a preference elicited before any interaction.

A controlled follow-up could distinguish these explanations by varying independently assessed attributes and concrete properties across comparable prompts, then eliciting abstract priorities and concrete preferences for the same dimensions before evaluation. Matching content and timing would isolate differences between elicitation formats. Manipulating psychological distance would test the construal account directly, while collecting more judgments per participant would support precise estimates of individual feature weights and a direct test of each report's incremental predictive value.

The scope is a single dataset of late-2023 models with capped response length, evaluated by crowdworkers in short sessions. Priorities were measured once, leaving their stability over time untested. Evidence for stable feature weights rests primarily on the REML models: simulation checks covered four features, the length test was somewhat liberal under participant-specific residual variances, and the moment-estimator interval for length included zero. Prediction intervals describe uncertainty across participants conditional on the fitted predictions and are unadjusted for the comparisons among information sources. The small gains therefore characterize the tested estimators and available histories. Longer interactions, repeated elicitation and newer models would establish how this pattern changes as personal preferences become more precisely estimable.

\section{Conclusion}

Stated preferences convey useful information about human evaluation of AI responses, and the information they convey depends on what is measured. In PRISM, relative priorities aligned with aggregate attribute profiles, including their attribute-specific components, while mean alignment with designed text proxies was near zero. Requests about length tracked the weight attached to length beyond those relative priorities, and reported reasons tracked perceived attributes. With short histories, stable differences in feature weights coexisted with poorly reproducible personal estimates and small average ranking gains from personal information. The central implication is to evaluate preference reports against their intended use: explaining perceived attributes, estimating concrete feature weights or predicting an individual's next choice.

\bibliographystyle{plainnat}
\bibliography{references}

@article{nisbett1977telling,
  title = {Telling more than we can know: Verbal reports on mental processes},
  author = {Nisbett, Richard E. and Wilson, Timothy D.},
  year = {1977},
  doi = {10.1037/0033-295X.84.3.231},
  journal = {Psychological Review},
  volume = {84},
  number = {3},
  pages = {231--259}
}

@article{nisbett1977halo,
  title = {The halo effect: Evidence for unconscious alteration of judgments},
  author = {Nisbett, Richard E. and Wilson, Timothy D.},
  year = {1977},
  doi = {10.1037/0022-3514.35.4.250},
  journal = {Journal of Personality and Social Psychology},
  volume = {35},
  number = {4},
  pages = {250--256}
}

@article{slovic1971comparison,
  title = {Comparison of {Bayesian} and regression approaches to the study of information processing in judgment},
  author = {Slovic, Paul and Lichtenstein, Sarah},
  year = {1971},
  doi = {10.1016/0030-5073(71)90033-X},
  journal = {Organizational Behavior and Human Performance},
  volume = {6},
  number = {6},
  pages = {649--744}
}

@article{slovic1995construction,
  title = {The construction of preference},
  author = {Slovic, Paul},
  year = {1995},
  doi = {10.1037/0003-066X.50.5.364},
  journal = {American Psychologist},
  volume = {50},
  number = {5},
  pages = {364--371}
}

@article{reilly1992assessment,
  title = {The assessment of self-insight in judgment policies},
  author = {Reilly, Barbara A and Doherty, Michael E},
  year = {1992},
  doi = {10.1016/0749-5978(92)90067-H},
  journal = {Organizational Behavior and Human Decision Processes},
  volume = {53},
  number = {3},
  pages = {285--309}
}

@article{karelaia2008determinants,
  title = {Determinants of linear judgment: A meta-analysis of lens model studies},
  author = {Karelaia, Natalia and Hogarth, Robin M.},
  year = {2008},
  doi = {10.1037/0033-2909.134.3.404},
  journal = {Psychological Bulletin},
  volume = {134},
  number = {3},
  pages = {404--426}
}

@article{karren2002review,
  title = {A Review and Analysis of the Policy-Capturing Methodology in Organizational Research: Guidelines for Research and Practice},
  author = {Karren, Ronald J. and Barringer, Melissa Woodard},
  year = {2002},
  doi = {10.1177/109442802237115},
  journal = {Organizational Research Methods},
  volume = {5},
  number = {4},
  pages = {337--361}
}

@article{aimansmith2002conducting,
  title = {Conducting Studies of Decision Making in Organizational Contexts: A Tutorial for Policy-Capturing and Other Regression-Based Techniques},
  author = {Aiman-Smith, Lynda and Scullen, Steven E. and Barr, Steve H.},
  year = {2002},
  doi = {10.1177/109442802237117},
  journal = {Organizational Research Methods},
  volume = {5},
  number = {4},
  pages = {388--414}
}

@article{alter2009uniting,
  title = {Uniting the Tribes of Fluency to Form a Metacognitive Nation},
  author = {Alter, Adam L. and Oppenheimer, Daniel M.},
  year = {2009},
  doi = {10.1177/1088868309341564},
  journal = {Personality and Social Psychology Review},
  volume = {13},
  number = {3},
  pages = {219--235}
}

@article{kokolakis2017privacy,
  title = {Privacy attitudes and privacy behaviour: A review of current research on the privacy paradox phenomenon},
  author = {Kokolakis, Spyros},
  year = {2017},
  doi = {10.1016/j.cose.2015.07.002},
  journal = {Computers \& Security},
  volume = {64},
  pages = {122--134}
}

@article{barth2017privacy,
  title = {The privacy paradox -- {Investigating} discrepancies between expressed privacy concerns and actual online behavior -- {A} systematic literature review},
  author = {Barth, Susanne and de Jong, Menno D.T.},
  year = {2017},
  doi = {10.1016/j.tele.2017.04.013},
  journal = {Telematics and Informatics},
  volume = {34},
  number = {7},
  pages = {1038--1058}
}

@article{thorndike1920constant,
  title = {A constant error in psychological ratings},
  author = {Thorndike, E.L.},
  year = {1920},
  doi = {10.1037/h0071663},
  journal = {Journal of Applied Psychology},
  volume = {4},
  number = {1},
  pages = {25--29}
}

@article{johansson2005failure,
  title = {Failure to Detect Mismatches Between Intention and Outcome in a Simple Decision Task},
  author = {Johansson, Petter and Hall, Lars and Sikström, Sverker and Olsson, Andreas},
  year = {2005},
  doi = {10.1126/science.1111709},
  journal = {Science},
  volume = {310},
  number = {5745},
  pages = {116--119}
}

@article{mccarthy2010mtld,
  title = {{MTLD}, {vocd-D}, and {HD-D}: A validation study of sophisticated approaches to lexical diversity assessment},
  author = {McCarthy, Philip M. and Jarvis, Scott},
  year = {2010},
  doi = {10.3758/BRM.42.2.381},
  journal = {Behavior Research Methods},
  volume = {42},
  number = {2},
  pages = {381--392}
}

@article{flesch1948new,
  title = {A new readability yardstick},
  author = {Flesch, Rudolph},
  year = {1948},
  doi = {10.1037/h0057532},
  journal = {Journal of Applied Psychology},
  volume = {32},
  number = {3},
  pages = {221--233}
}

@article{cameron2015practitioner,
  title = {A Practitioner's Guide to Cluster-Robust Inference},
  author = {Cameron, A. Colin and Miller, Douglas L.},
  year = {2015},
  doi = {10.3368/jhr.50.2.317},
  journal = {Journal of Human Resources},
  volume = {50},
  number = {2},
  pages = {317--372}
}

@article{jurek2005proper,
  title = {Proper interpretation of non-differential misclassification effects: expectations vs observations},
  author = {Jurek, Anne M and Greenland, Sander and Maldonado, George and Church, Timothy R},
  year = {2005},
  doi = {10.1093/ije/dyi060},
  journal = {International Journal of Epidemiology},
  volume = {34},
  number = {3},
  pages = {680--687}
}

@article{kirk2024benefits,
  title = {The benefits, risks and bounds of personalizing the alignment of large language models to individuals},
  author = {Kirk, Hannah Rose and Vidgen, Bertie and Röttger, Paul and Hale, Scott A.},
  year = {2024},
  doi = {10.1038/s42256-024-00820-y},
  journal = {Nature Machine Intelligence},
  volume = {6},
  number = {4},
  pages = {383--392}
}

@inproceedings{zamfirescu2023johnny,
  title = {Why {Johnny} Can't Prompt: How Non-{AI} Experts Try (and Fail) to Design {LLM} Prompts},
  author = {Zamfirescu-Pereira, J.D. and Wong, Richmond Y. and Hartmann, Bjoern and Yang, Qian},
  year = {2023},
  doi = {10.1145/3544548.3581388},
  booktitle = {Proceedings of the 2023 CHI Conference on Human Factors in Computing Systems},
  pages = {1--21},
  publisher = {ACM}
}

@inproceedings{subramonyam2024bridging,
  title = {Bridging the Gulf of Envisioning: Cognitive Challenges in Prompt Based Interactions with {LLMs}},
  author = {Subramonyam, Hari and Pea, Roy and Pondoc, Christopher and Agrawala, Maneesh and Seifert, Colleen},
  year = {2024},
  doi = {10.1145/3613904.3642754},
  booktitle = {Proceedings of the CHI Conference on Human Factors in Computing Systems},
  pages = {1--19},
  publisher = {ACM}
}

@article{hutto2014vader,
  title = {{VADER}: A Parsimonious Rule-Based Model for Sentiment Analysis of Social Media Text},
  author = {Hutto, C. J. and Gilbert, Eric},
  year = {2014},
  doi = {10.1609/icwsm.v8i1.14550},
  journal = {Proceedings of the International AAAI Conference on Web and Social Media},
  volume = {8},
  number = {1},
  pages = {216--225}
}

@article{trope2010construal,
  title = {Construal-level theory of psychological distance},
  author = {Trope, Yaacov and Liberman, Nira},
  year = {2010},
  doi = {10.1037/a0018963},
  journal = {Psychological Review},
  volume = {117},
  number = {2},
  pages = {440--463}
}

@article{hammond1955probabilistic,
  title = {Probabilistic functioning and the clinical method},
  author = {Hammond, Kenneth R.},
  year = {1955},
  doi = {10.1037/h0046845},
  journal = {Psychological Review},
  volume = {62},
  number = {4},
  pages = {255--262}
}

@inproceedings{kirk2024prism,
  author = {Kirk, Hannah Rose and Whitefield, Alexander and Röttger, Paul and Bean, Andrew and Margatina, Katerina and Ciro, Juan and Mosquera, Rafael and Bartolo, Max and Williams, Adina and He, He and Vidgen, Bertie and Hale, Scott A.},
  booktitle = {Advances in Neural Information Processing Systems},
  doi = {10.52202/079017-3342},
  pages = {105236--105344},
  publisher = {Curran Associates, Inc.},
  title = {The {PRISM} Alignment Dataset: What Participatory, Representative and Individualised Human Feedback Reveals About the Subjective and Multicultural Alignment of Large Language Models},
  volume = {37},
  year = {2024}
}

@article{kirk2026prismx,
  title = {{PRISM-X}: Experiments on Personalised Fine-Tuning with Human and Simulated Users},
  author = {Kirk, Hannah Rose and {Liu Leqi} and Zeng, Fanzhi and Davidson, Henry and Vidgen, Bertie and Summerfield, Christopher and Hale, Scott A.},
  year = {2026},
  eprint = {2605.13307},
  archiveprefix = {arXiv},
  journal = {arXiv preprint arXiv:2605.13307}
}

@article{chen2026synthetic,
  title = {When Synthetic Users Fail: A Cross-Domain Benchmark of {LLM}-Simulated Human Survey Responses},
  author = {Chen, Zihan and Zhu, Di and Zheng, Lei Nico},
  year = {2026},
  eprint = {2607.26348},
  archiveprefix = {arXiv},
  journal = {arXiv preprint arXiv:2607.26348}
}

@inproceedings{ghafouri2026rlhf,
  title = {Position: {RLHF} May Not Reflect Genuine Preferences},
  author = {Ghafouri, Bijean and Choi, Eun Cheol and Dey, Priyanka and Ferrara, Emilio},
  booktitle = {Proceedings of the 43rd International Conference on Machine Learning},
  pages = {170283--170322},
  year = {2026},
  volume = {306},
  series = {Proceedings of Machine Learning Research},
  publisher = {PMLR},
  url = {https://proceedings.mlr.press/v306/ghafouri26a.html}
}

@inproceedings{hosking2024human,
  author = {Hosking, Tom and Blunsom, Phil and Bartolo, Max},
  booktitle = {International Conference on Learning Representations},
  title = {Human Feedback is not Gold Standard},
  url = {https://proceedings.iclr.cc/paper_files/paper/2024/file/f719b8ca81221f380836e573d479c223-Paper-Conference.pdf},
  year = {2024}
}

@inproceedings{singhal2023long,
  title = {A Long Way to Go: Investigating Length Correlations in {RLHF}},
  author = {Singhal, Prasann and Goyal, Tanya and Xu, Jiacheng and Durrett, Greg},
  year = {2024},
  url = {https://openreview.net/forum?id=G8LaO1P0xv},
  booktitle = {First Conference on Language Modeling}
}

@inproceedings{sharma2024towards,
  author = {Sharma, Mrinank and Tong, Meg and Korbak, Tomasz and Duvenaud, David and Askell, Amanda and Bowman, Samuel R. and Cheng, Newton and Durmus, Esin and Hatfield-Dodds, Zac and Johnston, Scott R. and Kravec, Shauna and Maxwell, Timothy and McCandlish, Sam and Ndousse, Kamal and Rausch, Oliver and Schiefer, Nicholas and Yan, Da and Zhang, Miranda and Perez, Ethan},
  booktitle = {International Conference on Learning Representations},
  title = {Towards Understanding Sycophancy in Language Models},
  url = {https://proceedings.iclr.cc/paper_files/paper/2024/file/0105f7972202c1d4fb817da9f21a9663-Paper-Conference.pdf},
  year = {2024}
}

@inproceedings{ouyang2022training,
  author = {Ouyang, Long and Wu, Jeff and Jiang, Xu and Almeida, Diogo and Wainwright, Carroll L. and Mishkin, Pamela and Zhang, Chong and Agarwal, Sandhini and Slama, Katarina and Ray, Alex and Schulman, John and Hilton, Jacob and Kelton, Fraser and Miller, Luke and Simens, Maddie and Askell, Amanda and Welinder, Peter and Christiano, Paul and Leike, Jan and Lowe, Ryan},
  booktitle = {Advances in Neural Information Processing Systems},
  doi = {10.52202/068431-2011},
  pages = {27730--27744},
  publisher = {Curran Associates, Inc.},
  title = {Training language models to follow instructions with human feedback},
  volume = {35},
  year = {2022}
}

@article{casper2023open,
  title = {Open Problems and Fundamental Limitations of Reinforcement Learning from Human Feedback},
  author = {Casper, Stephen and Davies, Xander and Shi, Claudia and Gilbert, Thomas Krendl and Scheurer, Jérémy and Rando, Javier and Freedman, Rachel and Korbak, Tomasz and Lindner, David and Freire, Pedro and Wang, Tony and Marks, Samuel and Segerie, Charbel-Raphaël and Carroll, Micah and Peng, Andi and Christoffersen, Phillip and Damani, Mehul and Slocum, Stewart and Anwar, Usman and Siththaranjan, Anand and Nadeau, Max and Michaud, Eric J. and Pfau, Jacob and Krasheninnikov, Dmitrii and Chen, Xin and Langosco, Lauro and Hase, Peter and Bıyık, Erdem and Dragan, Anca and Krueger, David and Sadigh, Dorsa and Hadfield-Menell, Dylan},
  year = {2023},
  url = {https://openreview.net/forum?id=bx24KpJ4Eb},
  journal = {Transactions on Machine Learning Research}
}

@inproceedings{sorensen2024roadmap,
  title = {Position: A Roadmap to Pluralistic Alignment},
  author = {Sorensen, Taylor and Moore, Jared and Fisher, Jillian and Gordon, Mitchell L and Mireshghallah, Niloofar and Rytting, Christopher Michael and Ye, Andre and Jiang, Liwei and Lu, Ximing and Dziri, Nouha and Althoff, Tim and Choi, Yejin},
  booktitle = {Proceedings of the 41st International Conference on Machine Learning},
  pages = {46280--46302},
  year = {2024},
  volume = {235},
  series = {Proceedings of Machine Learning Research},
  publisher = {PMLR},
  url = {https://proceedings.mlr.press/v235/sorensen24a.html}
}

@inproceedings{li2024dissecting,
  title = {Dissecting Human and {LLM} Preferences},
  author = {Li, Junlong and Zhou, Fan and Sun, Shichao and Zhang, Yikai and Zhao, Hai and Liu, Pengfei},
  booktitle = {Proceedings of the 62nd Annual Meeting of the {A}ssociation for {C}omputational {L}inguistics (Volume 1: Long Papers)},
  year = {2024},
  publisher = {Association for Computational Linguistics},
  doi = {10.18653/v1/2024.acl-long.99},
  pages = {1790--1811}
}

@inproceedings{reimers2019sentence,
  title = {{Sentence-BERT}: Sentence Embeddings using {Siamese} {BERT}-Networks},
  author = {Reimers, Nils and Gurevych, Iryna},
  booktitle = {Proceedings of the 2019 Conference on Empirical Methods in Natural Language Processing and the 9th International Joint Conference on Natural Language Processing ({EMNLP}-{IJCNLP})},
  year = {2019},
  publisher = {Association for Computational Linguistics},
  doi = {10.18653/v1/D19-1410},
  pages = {3982--3992}
}

@inproceedings{zheng2023judging,
  author = {Zheng, Lianmin and Chiang, Wei-Lin and Sheng, Ying and Zhuang, Siyuan and Wu, Zhanghao and Zhuang, Yonghao and Lin, Zi and Li, Zhuohan and Li, Dacheng and Xing, Eric P. and Zhang, Hao and Gonzalez, Joseph E. and Stoica, Ion},
  booktitle = {Advances in Neural Information Processing Systems},
  doi = {10.52202/075280-2020},
  pages = {46595--46623},
  publisher = {Curran Associates, Inc.},
  title = {Judging {LLM}-as-a-Judge with {MT-Bench} and {Chatbot Arena}},
  volume = {36},
  year = {2023}
}

@article{dubois2024length,
  title = {Length-Controlled {AlpacaEval}: A Simple Way to Debias Automatic Evaluators},
  author = {Dubois, Yann and Galambosi, Balázs and Liang, Percy and Hashimoto, Tatsunori B.},
  year = {2024},
  eprint = {2404.04475},
  archiveprefix = {arXiv},
  journal = {arXiv preprint arXiv:2404.04475}
}

@inproceedings{chiang2024chatbot,
  title = {{Chatbot Arena}: An Open Platform for Evaluating {LLMs} by Human Preference},
  author = {Chiang, Wei-Lin and Zheng, Lianmin and Sheng, Ying and Angelopoulos, Anastasios Nikolas and Li, Tianle and Li, Dacheng and Zhu, Banghua and Zhang, Hao and Jordan, Michael and Gonzalez, Joseph E. and Stoica, Ion},
  booktitle = {Proceedings of the 41st International Conference on Machine Learning},
  pages = {8359--8388},
  year = {2024},
  volume = {235},
  series = {Proceedings of Machine Learning Research},
  publisher = {PMLR},
  url = {https://proceedings.mlr.press/v235/chiang24b.html}
}

@inproceedings{santurkar2023whose,
  title = {Whose Opinions Do Language Models Reflect?},
  author = {Santurkar, Shibani and Durmus, Esin and Ladhak, Faisal and Lee, Cinoo and Liang, Percy and Hashimoto, Tatsunori},
  booktitle = {Proceedings of the 40th International Conference on Machine Learning},
  pages = {29971--30004},
  year = {2023},
  volume = {202},
  series = {Proceedings of Machine Learning Research},
  publisher = {PMLR},
  url = {https://proceedings.mlr.press/v202/santurkar23a.html}
}

@inproceedings{poddar2024personalizing,
  author = {Poddar, Sriyash and Wan, Yanming and Ivison, Hamish and Gupta, Abhishek and Jaques, Natasha},
  booktitle = {Advances in Neural Information Processing Systems},
  doi = {10.52202/079017-1664},
  pages = {52516--52544},
  publisher = {Curran Associates, Inc.},
  title = {Personalizing Reinforcement Learning from Human Feedback with Variational Preference Learning},
  volume = {37},
  year = {2024}
}

@inproceedings{ryan2025synthesizeme,
  title = {{SynthesizeMe}! {Inducing} Persona-Guided Prompts for Personalized Reward Models in {LLMs}},
  author = {Ryan, Michael J. and Shaikh, Omar and Bhagirath, Aditri and Frees, Daniel and Held, William and Yang, Diyi},
  booktitle = {Proceedings of the 63rd Annual Meeting of the {A}ssociation for {C}omputational {L}inguistics (Volume 1: Long Papers)},
  year = {2025},
  publisher = {Association for Computational Linguistics},
  doi = {10.18653/v1/2025.acl-long.397},
  pages = {8045--8078}
}

@article{keswani2025moral,
  title = {Moral Change or Noise? {On} Problems of Aligning {AI} with Temporally Unstable Human Feedback},
  author = {Keswani, Vijay and Cousins, Cyrus and Nguyen, Breanna and Conitzer, Vincent and Heidari, Hoda and Borg, Jana Schaich and Sinnott-Armstrong, Walter},
  year = {2026},
  journal = {Proceedings of the AAAI Conference on Artificial Intelligence},
  volume = {40},
  number = {44},
  pages = {37501--37509},
  doi = {10.1609/aaai.v40i44.41083}
}

@article{chung2025literarytaste,
  title = {{LiteraryTaste}: A Preference Dataset for Creative Writing Personalization},
  author = {Chung, John Joon Young and Padmakumar, Vishakh and Roemmele, Melissa and Wang, Yi and Sun, Yuqian and Wang, Tiffany and Almeda, Shm Garanganao and Halperin, Brett A. and Lu, Yuwen and Kreminski, Max},
  year = {2025},
  eprint = {2511.09310},
  archiveprefix = {arXiv},
  journal = {arXiv preprint arXiv:2511.09310}
}

\appendix

\section{Lexicons and feature details}
\label{app:lexicons}

All markers were matched case-insensitively on word boundaries. Viewpoint phrases: on the other hand; some people, argue, believe, think, may, might, would, say; others argue, believe, think, may, might, would, say, feel; perspective(s); viewpoint(s); point(s) of view; different views, opinions, beliefs, cultures, people; opinions vary or differ; while some; proponents; critics; supporters; opponents; both sides; debate(d); controversial; depends on; subjective; worldview(s); multiple views, factors, perspectives. Caution phrases: consult a, with, your; seek professional, medical, legal, help, advice, support; professional help, advice, support; please be careful, note, remember, consider; be careful; caution; safety; safely; risk(s); harm(ful); dangerous; it is important to note, remember, consider, be, seek; not a substitute, medical, legal or financial advice; emergency; hotline; responsibly; ethical(ly); respectful(ly). Refusal and disclaimer phrases: as an AI; as a language model, AI language model, large language model, chatbot, machine, virtual, conversational, digital or helpful assistant; I am (just) an AI; I cannot, can not, am unable to, am not able to, will not, followed by provide, help, assist, answer, give, share, express, offer, engage, comment, make, form or have; I am sorry, but; I do not have personal, opinions, beliefs, feelings, emotions. Validation phrases: great, good, excellent, interesting, wonderful or fantastic question; what a great, wonderful, lovely, fantastic; I (completely, totally, fully) understand; I am (so) glad; that is great, wonderful, amazing, fantastic; you are right, welcome; absolutely; thank you for sharing, asking, your; I appreciate; I am happy to; it is great or wonderful that you. Coherence defects: no terminal punctuation; a line beginning with a speaker label such as Reply, User or Human followed by a colon; more than 10\% repeated word trigrams. Informational density, lexical diversity and readability were winsorized at the 1st and 99th percentiles; lexical diversity is undefined below ten tokens (\CreMissing\% of responses), and these responses were set to their conversation mean with a missing-value indicator.

Free-text feedback codes. Feedback was lowercased and split into clauses at sentence punctuation and at contrastive conjunctions (but, however, although, though, yet). A match was discarded when a negator (not, never, no, without, nothing, hardly or a negated auxiliary) occurred in the four preceding words of its clause or inside a modal request. Requests for shorter responses: statements of excess (too long, lengthy, wordy, verbose or detailed; too much text, information or detail; long-winded, waffling, rambling, padding; more detail than needed or asked for); modal requests for shorter, briefer, more concise, condensed or more compact responses (should, would, could, might, wish, want, prefer, hope or need within a few words); statements of preferring short or concise answers; keep it short, brief or to the point; less text or fewer words. Such matches were discarded in clauses about the participant's own prompts, about truncation or about response latency, and bare descriptive words in praising clauses were not counted. Requests for longer responses: too short, brief, vague, general, generic, simple, basic or superficial; modal or need statements followed by more detail, depth, information, examples or explanation; going into more or greater detail or depth; elaborating or expanding further; lack of detail, depth, information or substance; not detailed, thorough or informative enough; not going into detail or elaborating. Such matches were discarded in clauses about links, sources or citations, about the participant's own prompt or about truncation. Brevity praise: concise, succinct, brief, short and sweet or to the point used predicatively in a praising clause, excluding requests such as more concise or should be brief. Detail praise: detailed, thorough, in-depth, comprehensive or informative. Formatting: bullet points, numbered, format(ting), paragraph(s), list form, listed. Refusal complaints: refuse(d), would not or could not answer, give an opinion or take a side, evasive, dodging, sidestepping, non-answer, cop-out, avoided answering, outside praising clauses. Requests and praise are never counted in the same category. The validation codebook asked six yes/no questions about each text: whether the writer wants the responses to be shorter or more concise; wants them to be longer or more detailed; praises a response for being brief, concise or to the point; praises a response for being detailed, thorough or informative; mentions the formatting or layout of a response; and complains that a response refused, avoided or evaded answering. Each question came with definitions and examples, and the coder saw only the texts, in random order.

\section{Analysis plan: deviations and exploratory additions}
\label{app:deviations}

\textbf{D1.} The plan defined fluency as Flesch reading ease \citep{flesch1948new}. Inspection of response texts, before any rating was examined, showed frequent coherence failures: \DefTrunc\% of analysed responses end mid-sentence because of the generation cap, \DefLeak\% fabricate a continuation of the dialogue, and \DefRep\% repeat themselves. Coherence became the primary feature and readability the alternative. \textbf{D2.} The plan required each feature to vary within 80\% of conversations, a threshold that is mis-specified for sparse markers; we required variation in at least 1,000 conversations. \textbf{D3.} Heavy-tailed features were winsorized and undefined lexical diversity was imputed (Appendix~\ref{app:lexicons}). \textbf{D4.} The planned prediction baseline contained one person-specific feature (similarity to the participant's self-description); the main prediction analysis uses a strictly non-personal baseline, and the planned version is reported in Table~\ref{tab:prereg}. \textbf{Exploratory additions.} The following analyses were added after the plan and are reported as exploratory: the perceptual-correspondence analysis, with its feature-level version and bivariate associations; the lens composites and the comparison of representations, with separate and joint models, repeated half-splits, a participant bootstrap through the construction, Holm-corrected contrasts and the specificity and reliability checks; the coding of concrete feedback, its independent validation and its prospective test; the common model for the length weight; the context interactions; the experience contrasts; the REML random-slope models with their calibration, the moment estimator and the recovery simulation; and the full prediction protocol, with a common baseline and common folds, training-only scaling and tuning, symmetric pairwise boosting with a linear pairwise comparator, forward-only evaluation and secondary metrics. In the planned robustness checks, the specification without model fixed effects keeps the position effects, the alternative-feature specification includes the matching main effects, and the leave-one-model-out analysis recomputes the within-conversation transformations after each deletion. Joint tests of all seven ipsatized priorities against a single feature use a six-column basis, because the seven ipsatized scores of a participant sum to zero.

\section{Additional results}
\label{app:extra}

\begin{table}[h]\centering\caption{Sample flow.}\label{tab:flow}\small\begin{tabular}{lr}
\toprule
Step & $n$ \\
\midrule
Survey participants & 1,500 \\
Participants with at least one conversation & 1,396 \\
Conversations & 8,011 \\
First-turn rated responses & 30,049 \\
\quad dropped: fewer than three words & 670 \\
\quad dropped: conversation left with fewer than two responses & 32 \\
Analysed responses & 29,347 \\
Analysed conversations & 7,978 \\
Analysed participants & 1,396 \\
\quad of whom identical ratings on all seven stated items (excluded from ipsatized analyses) & 33 \\
Participants in ipsatized analyses & 1,363 \\
Participants with at least four analysed conversations & 1,309 \\
\bottomrule
\end{tabular}\end{table}

\begin{table}[h]\centering\caption{Share of N/A reasons by attribute and conversation type.}\label{tab:na}\small\begin{tabular}{lrrr}
\toprule
Attribute & Unguided (\%) & Values guided (\%) & Controversy guided (\%) \\
\midrule
Values & 21.4 & 4.8 & 6.6 \\
Creativity & 9.5 & 8.2 & 11.2 \\
Fluency & 1.1 & 0.6 & 0.8 \\
Factuality & 3.0 & 2.0 & 1.2 \\
Diversity & 19.0 & 7.9 & 4.8 \\
Safety & 11.2 & 6.2 & 6.1 \\
Helpfulness & 0.9 & 0.8 & 1.2 \\
\bottomrule
\end{tabular}\end{table}

\begin{table}[h]\centering\caption{Perceptual correspondence of each designed feature with its attribute: (1) partial and (2) bivariate within-participant associations of the chosen-minus-other contrast; (3) partial association of the chosen response's feature level, controlling for conversation type and prompt length; and the feature most strongly associated with each perceived attribute under each version (standardized coefficient in parentheses).}\label{tab:cvsens}\small\begin{tabular}{>{\raggedright\arraybackslash}p{1.6cm}>{\raggedright\arraybackslash}p{2.6cm}rrr>{\raggedright\arraybackslash}p{3.1cm}>{\raggedright\arraybackslash}p{3.1cm}}
\toprule
Perceived attribute & Designed feature & (1) & (2) & (3) & Strongest, contrast & Strongest, level \\
\midrule
Values & Similarity to own self-description & 0.02 & 0.03 & 0.07 & Refusal or AI-disclaimer language ($-$0.04) & Refusal or AI-disclaimer language ($-$0.08) \\
Creativity & Lexical diversity & 0.01 & 0.01 & 0.03 & Lists and headings (0.06) & Refusal or AI-disclaimer language ($-$0.11) \\
Fluency & Coherence (no truncation, leakage, repetition) & 0.09 & 0.07 & 0.11 & Coherence (no truncation, leakage, repetition) (0.09) & Coherence (no truncation, leakage, repetition) (0.11) \\
Factuality & Informational density & 0.01 & 0.02 & 0.00 & Length (0.09) & Length (0.16) \\
Diversity & Viewpoint markers & 0.08 & 0.09 & 0.12 & Length (0.09) & Length (0.15) \\
Safety & Caution markers & $-$0.01 & $-$0.01 & $-$0.01 & Coherence (no truncation, leakage, repetition) (0.04) & Coherence (no truncation, leakage, repetition) (0.05) \\
Helpfulness & Relevance to prompt & 0.01 & 0.02 & 0.07 & Lists and headings (0.07) & Refusal or AI-disclaimer language ($-$0.12) \\
\bottomrule
\end{tabular}\end{table}

\begin{table}[h]\centering\caption{Attribute-level moderation of the weights on text proxies by stated importance, for three representations of importance. Coefficient (SE); $^{\dagger}$ Holm-adjusted $p < 0.05$ within the column.}\label{tab:attlevel}\small\begin{tabular}{lrrr}
\toprule
Attribute (text proxy) & Ipsatized & Raw (z) & Rank \\
\midrule
Values & 0.02 (0.19) & 0.13 (0.17) & 0.09 (0.25) \\
Creativity & $-$0.24 (0.17) & $-$0.16 (0.16) & $-$0.18 (0.19) \\
Fluency & 0.13 (0.26) & 0.06 (0.18) & 0.25 (0.25) \\
Factuality & $-$0.30 (0.28) & $-$0.08 (0.18) & $-$0.36 (0.26) \\
Diversity & 0.10 (0.21) & 0.12 (0.18) & 0.03 (0.22) \\
Safety & 0.20 (0.17) & 0.19 (0.16) & 0.27 (0.18) \\
Helpfulness & 0.69 (0.30) & 0.50 (0.17)$^{\dagger}$ & 0.44 (0.28) \\
\midrule
Mean of seven & 0.09 (0.08) & 0.11 (0.06) & 0.08 (0.09) \\
Joint Wald $\chi^2(7)$, $p$ & 10.7, 0.15 & 13.1, 0.07 & 9.1, 0.24 \\
\bottomrule
\end{tabular}\end{table}

\begin{table}[h]\centering\caption{Moderation under three representations of the attributes: attribute-level coefficients from the joint model (SE) and summaries of the mean across specifications and resampling schemes.}\label{tab:levels}\small\begin{tabular}{>{\raggedright\arraybackslash}p{6.0cm}rrr}
\toprule
Attribute or summary & Text proxies & Lens composites & Aggregate profiles \\
\midrule
Values & 0.10 (0.20) & 0.19 (0.21) & 0.17 (0.29) \\
Creativity & $-$0.07 (0.17) & 0.33 (0.24) & 0.31 (0.30) \\
Fluency & $-$0.20 (0.31) & 0.28 (0.37) & 1.43 (0.34) \\
Factuality & $-$0.36 (0.28) & 0.15 (0.33) & 0.71 (0.34) \\
Diversity & $-$0.19 (0.27) & 0.35 (0.30) & 0.94 (0.28) \\
Safety & 0.32 (0.17) & 0.63 (0.20) & 0.88 (0.27) \\
Helpfulness & 0.40 (0.31) & 0.39 (0.35) & 1.50 (0.39) \\
\midrule
Mean of seven, joint model & 0.00 (0.09) & 0.33 (0.14) & 0.85 (0.23) \\
Mean of seven, separate models & 0.08 (0.08) & 0.42 (0.12) & 0.93 (0.23) \\
Joint, half-specific model and position effects & 0.00 (0.09) & 0.33 (0.13) & 0.86 (0.23) \\
50 random half-splits, 2.5th--97.5th percentile & [$-$0.05, 0.03] & [0.12, 0.36] & [0.59, 1.14] \\
Participant bootstrap through construction, 95\% CI & [$-$0.17, 0.15] & [$-$0.10, 0.56] & [0.15, 1.13] \\
\bottomrule
\end{tabular}\end{table}

\begin{table}[h]\centering\caption{Pairwise contrasts between representations (joint model), with Holm correction across the three contrasts and participant-bootstrap intervals through the construction of the representations.}\label{tab:levcontr}\small\begin{tabular}{lrrrrr}
\toprule
Contrast of mean moderation (joint model) & Estimate & SE & $p$ & Holm $p$ & Bootstrap 95\% CI \\
\midrule
Profiles minus proxies & 0.85 & 0.25 & $<$0.001 & 0.002 & [0.16, 1.12] \\
Profiles minus lens & 0.51 & 0.28 & 0.070 & 0.127 & [$-$0.18, 0.98] \\
Lens minus proxies & 0.33 & 0.18 & 0.064 & 0.127 & [$-$0.14, 0.66] \\
\bottomrule
\end{tabular}\end{table}

\begin{table}[h]\centering\caption{Attribute specificity of the profile result. Each row is a separate model with the profile main effects and model fixed effects; mismatched pairings interact each priority with the profile of another attribute.}\label{tab:profspec}\small\begin{tabular}{>{\raggedright\arraybackslash}p{8.4cm}rrr}
\toprule
Profile representation interacted with stated importance & Mean moderation & SE & $p$ \\
\midrule
Matched aggregate profiles & 0.93 & 0.23 & $<$0.001 \\
Attribute-specific deviations (common perceived quality removed) & 0.29 & 0.08 & $<$0.001 \\
Mismatched pairings, mean over six cyclic shifts & $-$0.06 & -- & -- \\
Mismatched pairings, largest of six cyclic shifts & 0.28 & -- & -- \\
\bottomrule
\end{tabular}\end{table}

\begin{table}[h]\centering\caption{Reliability of the representations: split-half correlations of model profiles and of their attribute-specific deviations across the two halves of raters, and held-out within-person correlations of lens composites with perceived performance.}\label{tab:levrel}\small\begin{tabular}{>{\raggedright\arraybackslash}p{5.6cm}rrr}
\toprule
Attribute & Profile split-half $r$ & Deviation split-half $r$ & Lens held-out $r$ \\
\midrule
Values & 0.61 & 0.59 & 0.060 \\
Creativity & 0.53 & 0.40 & 0.057 \\
Fluency & 0.83 & 0.41 & 0.077 \\
Factuality & 0.72 & 0.72 & 0.094 \\
Diversity & 0.62 & 0.44 & 0.117 \\
Safety & 0.35 & 0.41 & 0.013 \\
Helpfulness & 0.74 & 0.40 & 0.074 \\
\midrule
Common perceived quality (mean over attributes) & 0.69 & -- & -- \\
\bottomrule
\end{tabular}\end{table}

\begin{table}[h]\centering\caption{Aggregate attribute profiles: construction checks (single representation, ipsatized importance).}\label{tab:profiles}\small\begin{tabular}{lrrr}
\toprule
Aggregate attribute profile specification & Mean moderation & SE & $p$ \\
\midrule
Profiles, no model FE (planned secondary check) & 1.24 & 0.23 & $<$0.001 \\
Profiles + model FE & 1.19 & 0.23 & $<$0.001 \\
Echo-residualized profiles & 1.26 & 0.24 & $<$0.001 \\
Echo-residualized profiles + model FE & 1.19 & 0.24 & $<$0.001 \\
Disjoint-half profiles & 0.94 & 0.21 & $<$0.001 \\
Disjoint-half profiles + model FE & 0.93 & 0.21 & $<$0.001 \\
\bottomrule
\end{tabular}\end{table}

\begin{table}[h]\centering\caption{Context dependence of pooled feature weights (reference: unguided conversations, average prompt length).}\label{tab:context}\small\begin{tabular}{llrrr}
\toprule
Feature & Interaction & Coefficient & SE & $p$ \\
\midrule
Length & $\times$ controversy-guided & 1.68 & 0.41 & $<$0.001 \\
Length & $\times$ values-guided & 0.96 & 0.40 & 0.017 \\
Length & $\times$ prompt length (SD) & 1.38 & 0.18 & $<$0.001 \\
Lists and headings & $\times$ controversy-guided & $-$0.54 & 0.51 & 0.284 \\
Lists and headings & $\times$ values-guided & $-$0.50 & 0.32 & 0.119 \\
Lists and headings & $\times$ prompt length (SD) & 0.25 & 0.18 & 0.165 \\
Refusal or AI-disclaimer language & $\times$ controversy-guided & 1.18 & 0.42 & 0.005 \\
Refusal or AI-disclaimer language & $\times$ values-guided & 0.73 & 0.48 & 0.131 \\
Refusal or AI-disclaimer language & $\times$ prompt length (SD) & $-$0.06 & 0.19 & 0.767 \\
Caution markers & $\times$ controversy-guided & 0.41 & 0.42 & 0.327 \\
Caution markers & $\times$ values-guided & 0.01 & 0.43 & 0.978 \\
Caution markers & $\times$ prompt length (SD) & 0.49 & 0.16 & 0.002 \\
Relevance to prompt & $\times$ controversy-guided & $-$1.43 & 0.39 & $<$0.001 \\
Relevance to prompt & $\times$ values-guided & $-$0.60 & 0.41 & 0.140 \\
Relevance to prompt & $\times$ prompt length (SD) & $-$0.25 & 0.18 & 0.169 \\
\bottomrule
\end{tabular}\end{table}

\begin{table}[h]\centering\caption{Validation of the free-text codes against an independent coder on a fresh stratified sample of \AudItems{} texts: texts the rules flag in the pool, texts checked, precision with Wilson 95\% confidence interval, and recall projected from sampled unflagged texts with a stratified bootstrap interval.}\label{tab:audit}\small\begin{tabular}{>{\raggedright\arraybackslash}p{6.2cm}rrrr}
\toprule
Code & Flagged in pool & Checked & Precision [95\% CI] & Recall [95\% CI] \\
\midrule
Request for shorter responses & 22 & 22 & 0.77 [0.57, 0.90] & 0.22 [0.12, 0.56] \\
Request for longer or more detailed responses & 295 & 40 & 0.75 [0.60, 0.86] & 0.18 [0.11, 0.38] \\
Praise of brevity & 143 & 20 & 1.00 [0.84, 1.00] & -- \\
Praise of detail & 587 & 20 & 0.85 [0.64, 0.95] & -- \\
Formatting mention & 196 & 30 & 0.80 [0.63, 0.90] & -- \\
Complaint about refusal or evasion & 34 & 10 & 0.70 [0.40, 0.89] & -- \\
\bottomrule
\end{tabular}\end{table}

\begin{table}[h]\centering\caption{Concrete feedback and the weights on the matching features. Coefficient per unit of the feedback score (net share of conversations with requests for shorter rather than longer responses, or share with formatting mentions); per SD: per standard deviation of the score over analysed responses. Holm correction within each family; both the main and sensitivity families are exploratory.}\label{tab:concrete}\small\begin{tabular}{>{\raggedright\arraybackslash}p{1.5cm}>{\raggedright\arraybackslash}p{1.9cm}>{\raggedright\arraybackslash}p{2.8cm}>{\raggedright\arraybackslash}p{2.1cm}rrrrr}
\toprule
Family & Feature & Timing of the report & Codes & Coef. & SE & Per SD & $p$ & Holm $p$ \\
\midrule
Main & Length & Concurrent (same conversation) & requests & $-$0.39 & 0.63 & $-$0.10 & 0.542 & 0.542 \\
Main & Length & Other conversations & requests & $-$6.16 & 1.54 & $-$0.78 & $<$0.001 & $<$0.001 \\
Main & Length & Earlier conversations only & requests & $-$3.50 & 1.05 & $-$0.68 & $<$0.001 & 0.002 \\
Main & Lists and headings & Other conversations & formatting mentions & 2.93 & 1.73 & 0.24 & 0.090 & 0.181 \\
\midrule
Sensitivity & Length & Other conversations & requests and praise & $-$4.14 & 0.88 & $-$0.89 & $<$0.001 & $<$0.001 \\
Sensitivity & Length & Earlier conversations only & requests and praise & $-$2.26 & 0.60 & $-$0.73 & $<$0.001 & $<$0.001 \\
\bottomrule
\end{tabular}\end{table}

\begin{table}[h]\centering\caption{Common model for the length weight: concrete feedback from other conversations and the seven stated priorities entered together.}\label{tab:common}\small\begin{tabular}{lrrrr}
\toprule
Priorities block & Concrete feedback $\times$ length & $p$ & Priorities $\times$ length, Wald (df) & $p$ \\
\midrule
Raw priorities & $-$6.23 (1.54) & $<$0.001 & 20.0 (7) & 0.006 \\
Ipsatized priorities & $-$6.38 (1.53) & $<$0.001 & 9.0 (6) & 0.171 \\
\bottomrule
\end{tabular}\end{table}

\begin{table}[h]\centering\caption{Do the stated priorities jointly moderate the weights on non-elicited features? Ipsatized priorities use a six-column basis because the seven scores sum to zero.}\label{tab:attrconcrete}\small\begin{tabular}{llrrr}
\toprule
Non-elicited feature & Stated scale & Joint Wald $\chi^2$ & df & $p$ \\
\midrule
Length & ipsatized (6-column basis) & 8.3 & 6 & 0.216 \\
Length & raw & 20.0 & 7 & 0.006 \\
Refusal or AI-disclaimer language & ipsatized (6-column basis) & 10.9 & 6 & 0.093 \\
Refusal or AI-disclaimer language & raw & 13.3 & 7 & 0.065 \\
Lists and headings & ipsatized (6-column basis) & 9.5 & 6 & 0.149 \\
Lists and headings & raw & 9.6 & 7 & 0.215 \\
\bottomrule
\end{tabular}\end{table}

\begin{table}[h]\centering\caption{Exploratory pairings of stated importance with non-elicited features, overall and by conversation type; Holm correction within each stated scale.}\label{tab:crosslevel}\small\begin{tabular}{lllrrrr}
\toprule
Stated importance $\times$ feature & Context & Stated scale & $\gamma$ & SE & $p$ & Holm $p$ \\
\midrule
Safety $\times$ Refusal or AI-disclaimer language & all & ipsatized & 0.40 & 0.22 & 0.062 & 0.559 \\
Safety $\times$ Refusal or AI-disclaimer language & all & raw & 0.39 & 0.20 & 0.055 & 0.439 \\
Safety $\times$ Caution markers & all & ipsatized & 0.19 & 0.17 & 0.262 & 1.000 \\
Safety $\times$ Caution markers & all & raw & 0.20 & 0.16 & 0.218 & 1.000 \\
Safety $\times$ Refusal or AI-disclaimer language & controversy & ipsatized & 0.35 & 0.29 & 0.222 & 1.000 \\
Safety $\times$ Refusal or AI-disclaimer language & other & ipsatized & 0.44 & 0.29 & 0.125 & 1.000 \\
Safety $\times$ Refusal or AI-disclaimer language & controversy & raw & 0.38 & 0.28 & 0.176 & 1.000 \\
Safety $\times$ Refusal or AI-disclaimer language & other & raw & 0.38 & 0.26 & 0.146 & 1.000 \\
Safety $\times$ Caution markers & controversy & ipsatized & 0.19 & 0.26 & 0.477 & 1.000 \\
Safety $\times$ Caution markers & other & ipsatized & 0.19 & 0.23 & 0.392 & 1.000 \\
Safety $\times$ Caution markers & controversy & raw & 0.27 & 0.24 & 0.268 & 1.000 \\
Safety $\times$ Caution markers & other & raw & 0.13 & 0.21 & 0.520 & 1.000 \\
Factuality $\times$ Length & all & ipsatized & $-$0.03 & 0.36 & 0.939 & 1.000 \\
Factuality $\times$ Length & all & raw & 0.10 & 0.23 & 0.648 & 1.000 \\
Helpfulness $\times$ Refusal or AI-disclaimer language & all & ipsatized & 0.22 & 0.37 & 0.556 & 1.000 \\
Helpfulness $\times$ Refusal or AI-disclaimer language & all & raw & 0.19 & 0.19 & 0.327 & 1.000 \\
Helpfulness $\times$ Lists and headings & all & ipsatized & 0.35 & 0.29 & 0.229 & 1.000 \\
Helpfulness $\times$ Lists and headings & all & raw & 0.33 & 0.16 & 0.041 & 0.373 \\
\bottomrule
\end{tabular}\end{table}

\begin{table}[h]\centering\caption{Specificity of reasons: standardized associations with perceived performance on all seven attributes entered together (complete cases, demeaned within participant).}\label{tab:specificity}\small\begin{tabular}{lrrrr}
\toprule
Reason attribute & $n$ & Same-attribute perception & Mean cross-attribute & Largest cross-attribute \\
\midrule
Values & 6,034 & 0.49 & 0.01 & 0.11 \\
Creativity & 6,050 & 0.56 & 0.01 & 0.06 \\
Fluency & 6,127 & 0.31 & 0.07 & 0.17 \\
Factuality & 6,120 & 0.47 & 0.04 & 0.17 \\
Diversity & 6,049 & 0.58 & 0.00 & 0.04 \\
Safety & 6,045 & 0.20 & 0.05 & 0.10 \\
Helpfulness & 6,126 & 0.43 & 0.05 & 0.15 \\
\bottomrule
\end{tabular}\end{table}

\begin{table}[h]\centering\caption{Participant-level and conversation-level random slopes, REML (within-person standardized ratings unless stated). $p$: likelihood ratio for the participant component against the 50:50 $\chi^2$ mixture; context: model with shared context interactions; raw: raw rating points.}\label{tab:likelihood}\small\begin{tabular}{>{\raggedright\arraybackslash}p{4.4cm}rrrrrrr}
\toprule
Feature & $\hat\tau_{\mathrm{person}}$ & $\hat\tau_{\mathrm{conv}}$ & LR & $p$ & Holm $p$ & $\hat\tau_{\mathrm{person}}$, context & $\hat\tau_{\mathrm{person}}$, raw \\
\midrule
Similarity to own self-description & 0.067 & 0.105 & 4.0 & 0.023 & 0.162 & 0.067 & 1.79 \\
Lexical diversity & 0.032 & 0.001 & 0.5 & 0.248 & 0.994 & 0.035 & 0.96 \\
Coherence (no truncation, leakage, repetition) & 0.100 & 0.162 & 15.0 & $<$0.001 & $<$0.001 & 0.101 & 2.72 \\
Informational density & 0.085 & 0.124 & 7.8 & 0.003 & 0.021 & 0.080 & 2.42 \\
Viewpoint markers & 0.046 & 0.074 & 1.1 & 0.144 & 0.864 & 0.050 & 1.37 \\
Caution markers & 0.046 & 0.083 & 1.1 & 0.148 & 0.864 & 0.043 & 0.99 \\
Relevance to prompt & 0.002 & 0.156 & 0.0 & 0.500 & 1.000 & 0.001 & 1.90 \\
Length & 0.171 & 0.171 & 102.6 & $<$0.001 & $<$0.001 & 0.165 & 4.39 \\
Lists and headings & 0.076 & 0.001 & 15.9 & $<$0.001 & $<$0.001 & 0.077 & 1.97 \\
Positive sentiment & 0.035 & 0.140 & 0.3 & 0.294 & 0.994 & 0.028 & 0.64 \\
Validation phrases & 0.000 & 0.002 & 0.0 & 0.500 & 1.000 & 0.000 & 0.14 \\
Refusal or AI-disclaimer language & 0.121 & 0.131 & 29.9 & $<$0.001 & $<$0.001 & 0.119 & 3.43 \\
\bottomrule
\end{tabular}\end{table}

\begin{table}[h]\centering\caption{Calibration of the participant-component test by parametric bootstrap (within-person standardized ratings): number of simulated datasets, observed likelihood ratio, 95th percentile and maximum of the simulated likelihood ratios, and the rejection rate of the nominal 5\% test with its Monte Carlo SE.}\label{tab:remlcal}\small\begin{tabular}{>{\raggedright\arraybackslash}p{3.2cm}>{\raggedright\arraybackslash}p{3.6cm}rrrrr}
\toprule
Feature & Simulated scenario & $n$ & Observed LR & 95th pct. & Max. & Rejection (MC SE) \\
\midrule
Length & Null, homoscedastic & 100 & 102.6 & 2.77 & 5.73 & 0.06 (0.02) \\
Length & Null, person-specific residual SD & 100 & 102.6 & 3.65 & 7.78 & 0.09 (0.03) \\
Refusal or AI-disclaimer language & Null, homoscedastic & 100 & 29.9 & 1.59 & 3.67 & 0.03 (0.02) \\
Coherence (no truncation, leakage, repetition) & Null, homoscedastic & 100 & 15.0 & 1.68 & 4.98 & 0.02 (0.01) \\
Lists and headings & Null, homoscedastic & 100 & 15.9 & 2.97 & 5.38 & 0.07 (0.03) \\
Length & Person SD at its fitted value & 50 & 102.6 & -- & -- & 1.00 (0.00) \\
\bottomrule
\end{tabular}\end{table}

\begin{table}[h]\centering\caption{Moment estimates of person-stable slope variance (raw rating points per feature SD; participant-bootstrap interval of the 50-split average) and split-half correlations of person slopes: observed, and simulated with participant-level standard deviations set to zero, to their fitted values or to twice those values.}\label{tab:moment}\small\begin{tabular}{>{\raggedright\arraybackslash}p{3.8cm}rrrrrrr}
\toprule
Feature & Pooled & $\hat\tau^2_{\mathrm{person}}$ & 95\% CI & Observed $r$ & Sim., zero & Sim., fitted & Sim., 2$\times$ \\
\midrule
Similarity to own self-description & 0.83 & 6.3 & [0.7, 13.0] & 0.049 & 0.000 & 0.019 & 0.058 \\
Lexical diversity & $-$0.07 & $-$3.9 & [$-$12.1, 4.2] & $-$0.024 & 0.001 & 0.008 & 0.014 \\
Coherence (no truncation, leakage, repetition) & 1.58 & 12.8 & [1.6, 24.7] & 0.048 & 0.004 & 0.018 & 0.062 \\
Informational density & $-$0.51 & 8.3 & [$-$3.2, 21.6] & 0.031 & 0.000 & 0.020 & 0.061 \\
Viewpoint markers & 0.65 & 6.6 & [$-$3.6, 18.5] & 0.029 & 0.002 & 0.017 & 0.055 \\
Caution markers & 0.05 & 8.4 & [$-$3.0, 20.7] & 0.036 & $-$0.001 & 0.020 & 0.057 \\
Relevance to prompt & 3.69 & 1.8 & [$-$5.3, 9.5] & 0.011 & 0.003 & 0.015 & 0.052 \\
Length & 5.56 & 5.9 & [$-$10.2, 19.9] & 0.018 & 0.001 & 0.044 & 0.133 \\
Lists and headings & 1.93 & 9.9 & [$-$3.1, 22.8] & 0.036 & 0.002 & 0.025 & 0.071 \\
Positive sentiment & 0.19 & $-$4.6 & [$-$11.6, 3.9] & $-$0.026 & $-$0.002 & 0.006 & 0.010 \\
Validation phrases & 0.09 & 6.2 & [$-$9.8, 22.2] & 0.025 & 0.001 & 0.020 & 0.040 \\
Refusal or AI-disclaimer language & $-$2.14 & 4.5 & [$-$6.8, 15.4] & 0.017 & 0.001 & 0.024 & 0.085 \\
\bottomrule
\end{tabular}\end{table}

\begin{table}[h]\centering\caption{Prediction of held-out rankings: gains in pairwise accuracy over the pooled baseline (gradient-boosting sliders row: over a symmetric boosting baseline), top-choice gains, change in squared rating error for linear rating models, share of participants with higher accuracy, and range of the average gain over the three repetitions. Intervals are participant-bootstrap intervals, not corrected for multiple comparisons.}\label{tab:predict}\small\begin{tabular}{>{\raggedright\arraybackslash}p{5.1cm}rrrrrr}
\toprule
Information added & Gain & 95\% CI & Top choice & $\Delta$MSE & Better (\%) & Repeats \\
\midrule
Own free-text self-description (similarity) & $-$0.03 & [$-$0.11, 0.05] & 0.01 & $-$0.1 & 22 & [$-$0.10, 0.02] \\
Own free-text wishes for AI (similarity) & 0.06 & [$-$0.01, 0.13] & 0.26 & 0.0 & 22 & [0.00, 0.12] \\
Importance sliders $\times$ features & $-$0.23 & [$-$0.45, 0.00] & $-$0.51 & 3.1 & 39 & [$-$0.30, $-$0.18] \\
Importance sliders $\times$ aggregate attribute profiles & 0.02 & [$-$0.11, 0.17] & 0.25 & $-$1.3 & 33 & [$-$0.06, 0.11] \\
Stated-profile neighbours & $-$0.11 & [$-$0.33, 0.12] & $-$0.20 & 0.5 & 38 & [$-$0.29, 0.00] \\
Demographics and AI experience $\times$ features & $-$0.55 & [$-$0.85, $-$0.26] & $-$1.06 & 8.6 & 41 & [$-$0.90, $-$0.33] \\
Reported reasons (other conversations) $\times$ features & $-$0.17 & [$-$0.40, 0.07] & $-$0.65 & 0.7 & 40 & [$-$0.23, $-$0.13] \\
Concrete feedback (other conversations) $\times$ matching features & 0.13 & [0.02, 0.24] & 0.29 & $-$0.3 & 23 & [0.12, 0.15] \\
Concrete feedback and importance sliders together & $-$0.20 & [$-$0.44, 0.04] & $-$0.46 & 2.8 & 40 & [$-$0.24, $-$0.15] \\
Other conversations: person-specific feature weights & $-$0.02 & [$-$0.18, 0.14] & $-$0.17 & $-$2.1 & 33 & [$-$0.13, 0.09] \\
Other conversations: person-specific model taste & 0.13 & [$-$0.04, 0.30] & 0.08 & $-$0.9 & 35 & [0.10, 0.15] \\
Importance sliders in symmetric gradient boosting (vs. boosting baseline) & $-$0.07 & [$-$0.26, 0.13] & $-$0.03 & -- & 40 & [$-$0.13, $-$0.03] \\
\midrule
Pooled linear pairwise logistic model instead of the linear rating model & 0.35 & [0.12, 0.62] & 1.11 & -- & 44 & [0.26, 0.45] \\
Pooled symmetric gradient boosting instead of the linear rating model & 1.11 & [0.73, 1.52] & 1.84 & -- & 53 & [1.00, 1.23] \\
Pooled symmetric gradient boosting instead of the linear pairwise model & 0.76 & [0.43, 1.10] & 0.73 & -- & 50 & [0.67, 0.87] \\
\bottomrule
\end{tabular}\end{table}

\begin{table}[h]\centering\caption{Direct paired comparisons of information sources (difference in participant-level pairwise accuracy).}\label{tab:watchask}\small\begin{tabular}{lrr}
\toprule
Paired comparison & Difference (pp) & 95\% CI \\
\midrule
\multicolumn{3}{l}{\textbf{Watch minus ask}} \\
\quad Model taste minus sliders $\times$ features & 0.36 & [0.10, 0.63] \\
\quad Model taste minus sliders $\times$ profiles & 0.11 & [$-$0.09, 0.32] \\
\quad Model taste minus stated-profile neighbours & 0.24 & [$-$0.02, 0.50] \\
\quad Feature weights minus sliders $\times$ features & 0.21 & [$-$0.05, 0.46] \\
\quad Feature weights minus sliders $\times$ profiles & $-$0.04 & [$-$0.25, 0.15] \\
\quad Feature weights minus stated-profile neighbours & 0.09 & [$-$0.15, 0.32] \\
\multicolumn{3}{l}{\textbf{Concrete feedback versus sliders}} \\
\quad Concrete feedback minus sliders $\times$ features & 0.36 & [0.12, 0.60] \\
\quad Concrete feedback minus sliders $\times$ profiles & 0.11 & [$-$0.06, 0.26] \\
\quad Concrete feedback and sliders minus sliders $\times$ features & 0.03 & [$-$0.07, 0.13] \\
\quad Concrete feedback and sliders minus concrete feedback & $-$0.33 & [$-$0.57, $-$0.11] \\
\multicolumn{3}{l}{\textbf{Free-text wishes versus other sources}} \\
\quad Own wishes minus model taste & $-$0.08 & [$-$0.26, 0.12] \\
\quad Own wishes minus concrete feedback & $-$0.07 & [$-$0.19, 0.05] \\
\bottomrule
\end{tabular}\end{table}

\begin{table}[h]\centering\caption{Forward-only prediction: gain (pp) and 95\% interval by number of earlier conversations, and the linear trend over history.}\label{tab:chrono}\small\begin{tabular}{lrrrr}
\toprule
Earlier conversations & Feature weights & Model taste & Reasons & Concrete feedback \\
\midrule
1 & 0.00 [$-$0.30, 0.31] & 0.16 [$-$0.12, 0.39] & $-$0.18 [$-$0.96, 0.53] & 0.21 [$-$0.15, 0.59] \\
2 & 0.07 [$-$0.25, 0.41] & 0.01 [$-$0.33, 0.34] & $-$0.36 [$-$0.95, 0.25] & 0.02 [$-$0.29, 0.34] \\
3 & $-$0.33 [$-$0.67, $-$0.01] & $-$0.24 [$-$0.69, 0.17] & $-$0.47 [$-$1.10, 0.11] & 0.06 [$-$0.22, 0.34] \\
4 & 0.03 [$-$0.35, 0.39] & 0.07 [$-$0.44, 0.54] & 0.15 [$-$0.47, 0.78] & 0.12 [$-$0.14, 0.36] \\
5 or more & 0.04 [$-$0.36, 0.44] & 0.32 [$-$0.17, 0.79] & $-$0.03 [$-$0.65, 0.61] & 0.17 [$-$0.11, 0.44] \\
\midrule
Trend per earlier conversation & 0.00 [$-$0.10, 0.11] & 0.04 [$-$0.09, 0.16] & 0.08 [$-$0.13, 0.29] & 0.00 [$-$0.09, 0.10] \\
\bottomrule
\end{tabular}\end{table}

\begin{table}[h]\centering\caption{Planned prediction contest (baseline includes similarity to the participant's self-description).}\label{tab:prereg}\small\begin{tabular}{lrrr}
\toprule
Planned model & Accuracy (\%) & Gain (pp) & 95\% CI \\
\midrule
P0 pooled features + model + position & 65.59 & -- & -- \\
P1 + demographics $\times$ features & 65.27 & $-$0.32 & [$-$0.68, 0.04] \\
P2 + stated $\times$ features & 65.53 & $-$0.06 & [$-$0.32, 0.19] \\
P3 + reasons $\times$ features & 65.54 & $-$0.05 & [$-$0.36, 0.24] \\
P5 + person-specific weights (other conversations) & 65.68 & 0.09 & [$-$0.16, 0.33] \\
P6 + stated, demographics and reasons & 64.93 & $-$0.66 & [$-$1.07, $-$0.25] \\
\bottomrule
\end{tabular}\end{table}

\begin{table}[h]\centering\caption{Experience: direct contrasts in mean moderation (text proxies).}\label{tab:experience}\small\begin{tabular}{lrrr}
\toprule
Contrast in mean moderation & Difference & SE & $p$ \\
\midrule
Very familiar minus not familiar & $-$0.26 & 0.22 & 0.255 \\
Very familiar minus somewhat familiar & 0.01 & 0.16 & 0.968 \\
Weekly or more minus less than weekly & $-$0.08 & 0.16 & 0.630 \\
\bottomrule
\end{tabular}\end{table}

\begin{table}[h]\centering\caption{Stated importance by randomized display position (participant and attribute fixed effects).}\label{tab:order}\small\begin{tabular}{lrr}
\toprule
Display position & Difference vs position 1 (points) & SE \\
\midrule
2 & 0.03 & 0.64 \\
3 & $-$1.17 & 0.65 \\
4 & $-$1.15 & 0.64 \\
5 & $-$0.58 & 0.65 \\
6 & $-$1.06 & 0.65 \\
7 & $-$1.42 & 0.65 \\
8 & $-$0.42 & 0.67 \\
\bottomrule
\end{tabular}\end{table}

\end{document}